\documentclass[3p,times]{elsarticle}
\usepackage[utf8]{inputenc}
\usepackage[T1]{fontenc}
\usepackage{graphicx}
\usepackage{eurosym}
\usepackage{amsfonts}
\usepackage{amssymb}
\usepackage{amsmath}
\usepackage{booktabs}
\usepackage{xcolor}

\usepackage{graphicx}
\usepackage{graphicx,epsfig}
\usepackage[font={footnotesize,it}]{caption}
\usepackage[colorlinks=false,
linkcolor=red,
urlcolor=red,
citecolor=blue]{hyperref}

\usepackage{float}
\usepackage{color}
\usepackage{mathptmx}
\usepackage{anyfontsize}
\usepackage{t1enc}
\usepackage{amssymb}
\usepackage[figuresright]{rotating}
\begin{document}
	
	\begin{frontmatter}
	\title{A study of  anisotropic compact star with MIT bag EoS in  $f(\Re,T)$ modified theory }
		\author{H. D. Singh}
	\ead{drdipalisingh25@gmail.com}
	\author{J. Kumar}
	\ead{jitendark@gmail.com}

	\address{Department of Mathematics, Central University of Jharkhand, Ranchi-835205, India}
	
	\begin{abstract}
		In the current study, we investigated a specific model of anisotropic strange stars specially Her X-1, in the background of modified $f(\Re,T)$ gravity by choosing $f(\Re,T)=\Re+2\xi T$, where $\Re$ is Ricci scalar, $T$ is the trace of the energy- momentum tensor and $\xi$ is a coupling constant. To obtained the solution for the modified field equations, we apply Buchdahl metric to our equations. We consider the case, when the matter is governed by MIT bag model equation of state as $P_r=\frac{1}{3}(\rho-4B)$, where $B$ is bag constant. We calculate the values of unknown parameters using Schwarzschild interior space-time followed by choosing the appropriate values of parameter $\xi$, $K$, $\beta$ and also tabulated different values of bag constant for different $\xi$. We examine the physical validity of our model by performing tests such as energy conditions, equilibrium of the forces, the adiabatic index, redshift and some more. The observational results showed that the proposed $f(\Re,T)$ model satisfies all these tests and are quite acceptable. 	
	\end{abstract}
	
	
\end{frontmatter}
	\section{Introduction}\label{sec1}
		Einstein's general theory of relativity has been  continued to be very helpful in uncovering hidden mystery of nature. Investigating the structural properties of high density compact stars specially neutron stars has interested many researchers both from mathematics and physics society. Pulsar stars including high magnetic spinning stars were also considered as some of high density compact stars in realm of astrophysics. Because of lack of knowledge about these compact stars, it is believed that the material content of such stars are subatomic particles. From predicting mercury's orbit and deflection of starlight because of some massive gravitational objects to detecting gravitational waves and black holes in space, GR has been remarkably used in concluding the phenomena of Cosmic Microwave Background Radiation \cite{spergel}, Planck data \cite{ade}, $\wedge$ Cold Dark Matter \cite{ketov}.   Still this theory has been challenged by the presence of late-time acceleration of the universe and the presence of the dark matter \cite{riess}\cite{perlmutter}\cite{padmanabhan}. As a result of which, many gravitational modified theory have been proposed such as $f(\Re)$ theory where $\Re$ is the Ricci Scalar, $f(T,\tau)$ and where $T$ and $\tau$ are energy-momentum tensor trace and torsion scalar, Gauss-Bonnet theory, Cartan gravity, Rosen bimetric, etc.  At present, Harko and his collaborators \cite{harko} look forward and designed $f(\Re,T)$ gravity theory to work against the accelerated expansion of the universe and dark energy matter\cite{singh} in the framework of $f(\Re,T)$ gravity theory. Many studies have been done \cite{cosmol}\cite{sharif}\cite{das} in context of anisotropic charged and uncharged models. Pretel and et.al studied the  static configuration of compact stars \cite{pretel} with context of $f(\Re,T)$. The models were constructed for $f(\Re,T)=\Re+2f(T)$ and $f(\Re,T)=f_1(\Re)+f_2(T)$. Many authors, work in $f(\Re,T)$ of FLRW with modified Chaplygin gas \cite{nagpal}, Little Rip model \cite{houndjo} that explains present stages of our universe. Santos and Ferst \cite{santos} studied the problem of casuality in $f(\Re,T)$ modified gravity. Zubair and et.al \cite{zubair} studied static spherical symmetric of wormholes in framework of $f(\Re,T)$ gravity. Gravitational decoupling by minimal geometric deformation model was studied by Azmat abd Zubair \cite{azmat}. Shabani and Farhoudi \cite{shabani} uses dynamical procedure and found solutions for $f(\Re,T)$ theory by considering parameters like Hubble parameter and EoS parameter.
		\par Now, different approaches were made to find analytic solutions of GR equations. One is by embedding approach of four-dimensional spacetime. Such type of approach is successful in acquisition of exact models \cite{kuchowicz}. In numerous literatures, researchers showed solutions through embedding class-I and class-II, solutions are also obtained by assuming various metric functions \cite{ivanov}\cite{kohler}. Karmarkar condition has also been used in many model of compact stars in the frame of different modified theories \cite{prasad}. Kumar and et.al has used Buchdahl metrics function in the framework of $f(\Re,T)$ gravity \cite{kumar}. Zubair and Abbas \cite{abbas} studied interior model for compact stars in the framework of $f(\Re,T)$ gravity using Krori and Barua solution. Moraes \cite{moraes} studied stellar equilibrium in background of $f(\Re,T)$ theory by using hydrostatic equilibrium equation. 
		\par In search of more realistic stellar configuration the researchers need to be connected with the macroscopic properties of stars which is determined through observations of the microphysics. A new method of stellar structure modeling was born  for which standard approaches were made which includes assumptions of the metric function, energy density, anisotropic parameter, matter content which allows connection to physics , masses linked to redshift and compactness of stellar structure model. EoS is one such approach. It basically links the radial pressure to the density to microphysics linearly. Applying barotropic EoS basically helps in understanding complicated microphysics at a macroscale level. We consider EoS in the form $p_r=p_r(\rho)$ that reduces the difficulty in finding exact model of the modified field equations by single generating function \cite{maharaj}\cite{sharma}. The colour-flavour locked EoS \cite{rej} describes strange quark matter. It is also used in examining the surface tension of stars which restrict the bag constant and tangential pressure included with the EoS model. 
		\par For constructing a model, Negreiros and collaborators have modeled for compact objects consisting of matter which follows MIT bag modeling (EoS) on the surface of the compact stars\cite{weber}\cite{weber1}. In these paper, we have studied a new class of anisotropic general solutions to modified Einsteins general relativity field equations for relativistic compact stellar models by considering Buchdahl metrics within the framework of $f(\Re,T)$ gravity theory.And simplified the solutions using MIT bag EoS equations for the compact stars Her X-1. This paper is structured as follows: in Sec.2, the basic of MIT bag EoS is presented. In Sec.3, a general mathematical framework of our $f(\Re,T)$ gravity theory has been shown, in sec.4, anisotropic matter distributions in $f(\Re,T)$ gravity theory and their solutions of anisotropic Buchdahl metric are b discussed. In Sec.5, we have been represented the matching condition for interior space-time in the framework of $f(\Re,T)$ theory by smooth joining with exterior Schwarzchild geometry. In Sec.6, we have described various physical properties for present model analytically as well as graphically. In Sec.7, we concluded the detail discussion on the determined results and their implication on the various cosmological cases.  

\section{MIT bag EOS}\label{sec2}
		In study for compact objects, it is important to know that the principle of thermodynamic  in which variables are connected. It is described that the microphysics properties of compact stars are linked to the macroscopic properties by EoS, particularly their radii and masses. Now depending upon the nature of interaction, the models were categorized by threebroad types: 1. Non-relativistic potential models, 2. Relativistic theoretical field models and 3. Relativistic DBHF models \cite{maurya}. Futhermore, one can classify EoS in two classes : first, EoS have a pressure which vanishes when density tends to zero and second, self-bound EoS have a pressure which vanishes at significantly finite density whereas our MIT bag model EoS can be referred to be self bound EoS \cite{witten}.
		\par Now, compact stars are basically considered with strange quark matter which is the ultimate ground of state of matter. It basically describes that the inward and outward pressure of the bag helps in stabilizing the system \cite{alcock} \cite{zdunik}. So, the MIT bag EoS model is the simplest equation of state to study equilibrium configuration of such models. To clear the properties of compact stars with high densities, many considered EoS $p_r=p_r(\rho)$ as approximated linear function of energy density \cite{dey}\cite{cheng}\cite{rosinska}. Mathematically, MIT bag EoS model is given by
		\begin{equation}
			p_r=\frac{1}{3}(\rho-4B) \label{1}
		\end{equation}
		where B is the bag constant. 
		In original MIT bag model, the value of bag constant was $B=55 Mev/fm^3$. In recent articles, the range of $B$ are found to be from $60$ to $90 Mev/fm^3$ for strange quark stars \cite{burgio}\cite{rahaman}\cite{alaverdyan}. 
		\par However, a linear relationship guarantees the casuality condition. Although the relationship between radial pressure and energy density is complex, still the core of the object is approximately linear and the standard equation for a stellar structure is given by the following expression
		\begin{equation}
			p_r=\chi(\rho-\rho_s) \label{2}
		\end{equation}
		where $\chi$ is a positive constant. It is clear from the equation that for $r=R$, the radial pressure vanishes i.e $p_r=0$. This happens due to the energy density at zero pressure i.e $\rho(R)=\rho_s$, which is a non-vanishing quantity. In case, EoS obeys polytropic form then the density vanishes at the boundary $p_r=0$.
		
\section{Basic field equation in $f(\Re,T)$ gravity}\label{sec 3}
		In this section, we formulate the general $f(\Re,T)$ as in \cite{harko} gravity theory. While obtaining the Einstein's field equations, the Ricci scalar in Einstein-Hilbert action is given as
		\begin{eqnarray}
			S_{E}=\frac{1}{16}\int{\Re\sqrt{-g}}d^4 x\label{3}
		\end{eqnarray} 
		Now one can get $f(\Re,T)$ field equation from eqn.{\ref{3}}, if we choose $f(\Re,T)$  in place of Ricci scalar $\Re$. Thus, the full action for the modified theory of gravity is given by 
		\begin{eqnarray}
			S=\frac{1}{16}\int{f(\Re,T)\sqrt{-g}d^4 x}+\int{L_M \sqrt{-g}d^4 x}\label{4}
		\end{eqnarray} 
		where in the first term $ f(\Re,T) $ is  an arbitrary function of Ricci scalar $\Re$ with $g$ as the determinant of the metric tensor $g_{\mu\nu}$ and $T$ is the trace of the stress-energy tensor,$T_{\mu\nu}$. And in the second term,  $L_M$ is the matter Lagrangian density which is related to stress-momentum tensor and given as 
		\begin{eqnarray}
			T_{\mu\nu}=-\frac{2}{\sqrt{-g}} \frac{\delta(\sqrt{-g} L_M)}{\delta g^{\mu \nu}} \label{5}
		\end{eqnarray}
		with trace $T=g^{\mu\nu}T_{\mu\nu}$. Assuming that the Lagrangian density $L_M$ only depends on the metric tensor components $g_{\mu\nu}$, we obtain eqn.(\ref{5}) as  
		\begin{eqnarray}
			T_{\mu\nu}=g_{\mu\nu}L_M-\frac{2\partial( L_M)}{\partial g^{\mu \nu}} \label{6}
		\end{eqnarray}
		By varying eqn.(\ref{4}) with respect to $g_{\mu\nu}$, we obtained the field equation 
		\begin{align}
			(\Re_{\mu\nu}-\bigtriangledown_{\mu}\bigtriangledown_{\nu})f_{\Re}(\Re,T)+g_{\mu\nu}\diamondsuit  f_{\Re}(\Re,T)-\frac{1}{2}f(\Re,T)g_{\mu\nu} \nonumber\\
			=& 
			8\pi T_{\mu\nu}-f_{T}(\Re,T)(T_{\mu\nu}+\varPsi_{\mu\nu}) \label{7}
		\end{align}	
		where $f_\Re(\Re,T)=\frac{\partial f(\Re,T)}{\partial \Re}$ and $f_T(\Re,T)=\frac{\partial f(\Re,T)}{\partial T}$. The covariant derivative is denoted by $\bigtriangledown_{\mu}$ and the diamond operator $\diamondsuit$ is the D'Alambert operator which is defined as 	 
		\begin{eqnarray*}
			\diamondsuit=\frac{\partial_{\mu}(\sqrt{-g}g^{\mu\nu}\partial_{\nu})}{\sqrt{-g}} 
		\end{eqnarray*}
		
		\begin{eqnarray*}
			\varPsi_{\mu\nu}=g^{\alpha\beta}\frac{\delta T_{\alpha\beta}}{\delta g^{\mu\nu}}
		\end{eqnarray*}
		Now we performed the covariant derivative of eqn.(\ref{7}) by which we obtained the energy-momentum tensor as 
		\begin{eqnarray}
			\bigtriangledown^{\mu} T_{\mu\nu}=\frac{f_T(\Re,T)}{8\pi-f_T(\Re,T)} [ (T_{\mu\nu}+\varPsi_{\mu\nu})\bigtriangledown^{\mu}ln f_T(\Re,T)+\varPsi_{\mu\nu}\bigtriangledown^{\mu}-\frac{1}{2}g^{\mu\nu}\bigtriangledown^{\mu} T ]  \label{8}
		\end{eqnarray}
		Clearly from eqn.(\ref{8}) we can see that the stress-energy momentum tensor $T_{\mu\nu}$ in $f(\Re,T)$ theory does not follow the conservation law as in Einstein general relativity (GR) due to the presence of some non-minimal matter geometry coupling in its formulation. By using eqn.(\ref{6}), the tensor $\varPsi_{\mu\nu}$ is given by
		\begin{eqnarray}
			\varPsi_{\mu\nu}=-2T_{\mu\nu}+g_{\mu\nu}L_M -2g^{\alpha\beta}\frac{\partial^2 L_M}{\partial g^{\mu\nu}\partial g^{\alpha\beta}} \label{9}
		\end{eqnarray} 
		To find the field equations, we consider the interior of star to be filled with a perfect fluid source along with the energy-momentum tensor of the form
		\begin{eqnarray}
			T_{\mu\nu}=(\rho+p_t)u_{\mu}u_{\nu}-p_{t} g_{\mu\nu}+(p_{r}-p_{t}) g_{\mu\nu}\label{10}
		\end{eqnarray}   
		provided the four-velocity $u_{\nu}$ satisfied $u_{\mu}u^{\mu}=-1$ and $u_{\nu}\bigtriangledown^{\mu} u_{\nu}=0$, $p_{r}$ and $ p_{t} $ is the radial and tangential pressure respectively . Taking the matter Lagrangian as $L_M=-p$ where $p=\frac{1}{3}(p_r+2p_t)$ \cite{Ortiz1,Errehymy}, we obtain eqn.(\ref{9}) as 
		\begin{eqnarray}
			\varPsi_{\mu\nu}=-2T_{\mu\nu}-pg_{\mu\nu} \label{11}
		\end{eqnarray}
		In this paper, we assumed the function $f(\Re,T)$as $f(\Re,T)=\Re+2f(T)$ where $f(T)$ is the function of trace $(T)$ of the stress-energy tensor of matter. Here, we choose $f(T)=\xi T$ to determine the modified theory of gravity, where $\xi$ is a coupling constant. \\
		Using the expression in eqn.(\ref{7}), we obtain 
		\begin{eqnarray}
			G_{\mu\nu}=8\pi T_{\mu\nu}+\xi Tg_{\mu\nu}+2\xi(T_{\mu\nu}+pg_{\mu\nu}) \label{12}
		\end{eqnarray}
		When $f(\Re,T)\equiv \Re$, then eqn.(\ref{7}) reduces to Einstein field equations. By putting $f(\Re,T)=\Re+2\xi T$ and eqn.(\ref{11}) in eqn.(\ref{8}) we get
		\begin{eqnarray}
			\bigtriangledown^{\mu}T_{\mu\nu}=-\frac{\xi}{2(4\pi+\xi)}[g_{\mu\nu}\bigtriangledown^{\mu}T+2\bigtriangledown^{\mu}(pg_{\mu\nu})]\label{13}
		\end{eqnarray}
		Now, we can write eqn.{\ref{13}} in form of 
		\begin{eqnarray*}
			\bigtriangledown^{\mu}T^{eff}_{\mu\nu}=0
		\end{eqnarray*}
		So the above equation represents the conservation of the energy-momentum tensor $T^{eff}_{\mu\nu}$ for effective matter distribution. Thus, we can easily obtain the conservation equation in Einstein's gravity by putting $\xi=0$.
		For studying cosmological and astrophysical phenomenon, $f(\Re,T)$ has been widely considered because of its advantages in explaining such space cases.
		
\section{Field equation in $f(\Re,T)$ modified theory}\label{sec 4}
		We assumed the interior spacetime of the spherically symmetric stellar structure by the metric coordinates $(t,r,\theta, \phi)$
		\begin{equation}
			ds^2=e^{\nu(r)}dt^2-e^{\lambda(r)}dr^2-r^2(d\theta^2+sin^2d\phi^2)\label{14}
		\end{equation}	
		where $\nu$ and $\lambda$ are the metric potentials and are function of the radial coordinate, $r$.
		\par Now, by using eqns \ref{10}, \ref{12} and \ref{14}, we have the Einstein field equations for $f(\Re,T)$ gravity theory as
		\begin{eqnarray}
			8\pi \rho^{eff}=e^{-\lambda}\big(\frac{\lambda^{\prime}}{r}-\frac{1}{r^2}\big)+\frac{1}{r^2}=8\pi \rho+\frac{\xi}{3}(9\rho-p_r-2p_t) \label{15}\\
			8\pi p_r^{eff}=e^{-\lambda}\big(\frac{\nu^{\prime}}{r}+\frac{1}{r^2}\big)-\frac{1}{r^2}=8\pi p_r-\frac{\xi}{3}(3\rho-7p_r-2p_t) \label{16}\\
			8\pi p_t^{eff}=\frac{e^{-\lambda}}{2}\big(\nu^{\prime\prime}+\frac{1}{2}\nu^{\prime^2}+\frac{\nu^{\prime}-\lambda^{\prime}}{r}-\frac{1}{2}\nu^{\prime}\lambda^{\prime}\big)=8\pi p_t-\frac{\xi}{3}(3\rho-p_r-8p_t)\label{17} 								
		\end{eqnarray}
		where the prime($\prime$)denote the differentiation with respect,$r$.
		\par To solve eqns \ref{15}, \ref{16} and \ref{17}, we used Buchdahl solution which was generated through a assuming spherically symmetric fluid spheres of Einstein's equation. The most widely studied metric ansatz is Buchdahl metric 
		\begin{eqnarray}
			e^{\lambda}=\frac{K(1+Cr^2)}{K+Cr^2}, K=\frac{7}{4}\label{18}
		\end{eqnarray} 
		Where $K$ and $C$ are two parameters that are characterized by the geometry of the star. For complete solution of these equations, we introduce transformations 
    	\begin{eqnarray}
	   		e^{\nu}=Y^2 \label{19}
		\end{eqnarray}
	
		For this anisotropic matter distribution in modified theory, it is necessary that the anisotropic fluid must satisfy conservation equation as
		\begin{eqnarray}
			p_r^{\prime}+\frac{\nu^{\prime}}{2}(\rho+p_r)-\frac{2}{r}(p_r-p_t)=\frac{\xi(3\rho^{\prime}-p_r^{\prime}-2p_t^{\prime})}{6(4\pi+\xi)}\label{20}
		\end{eqnarray}
		 
		In another way, the energy density $(\rho)$, radial pressure $(p_r)$ and tangential pressure $(p_t)$ for anisotropic stellar model in $f(\Re,T)$ theory are given as
		\begin{eqnarray}
			\rho=\frac{8\pi\rho^{eff}}{(8\pi+4\xi)}+\frac{8\pi(3\rho^{eff}+p_r^{eff}+2p_t^{eff})\xi}{3(8\pi+2\xi)(8\pi+4\xi)}\label{21}\\
			p_r=\frac{8\pi p_r^{eff}}{(8\pi+2\xi)}+\frac{8\pi(3\rho^{eff}-p_r^{eff}-2p_t^{eff})\xi}{3(8\pi+2\xi)(8\pi+4\xi)}\label{22}\\
			p_t=\frac{8\pi p_t^{eff}}{(8\pi+2\xi)}+\frac{8\pi(3\rho^{eff}-p_r^{eff}-2p_t^{eff})\xi}{3(8\pi+2\xi)(8\pi+4\xi)}\label{23}
		\end{eqnarray}
		Where $\xi noteq -4\pi$ and $-2\pi$. For solving eqns \ref{21} - \ref{23}, we have assumed the equation of state(EoS) in the interior of the stellar structure to be governed by MIT bag model. For anisotropic fluid, we have considered the relation between the energy density and the radial pressure of the fluid by 
		\begin{eqnarray}
			\rho=3p_r+4B \label{24}
		\end{eqnarray} 
		 So, from eqn \ref{24}, we derive the well known MIT bag model EoS as 
		\begin{eqnarray}
			p_r=\frac{1}{3}(\rho-4B) \label{25}
		\end{eqnarray}
		where $B$ is the bag constant approximation between the range of the values $57 MeV/fm^3$ to $94 MeV/fm^3$.
		With introduction to MIT bag model, eqns \ref{15}, \ref{16} and \ref{17} can be written as
		\begin{eqnarray}
		    \rho^{eff}=\frac{3(3+Cr^2)}{56\pi(1+Cr^2)^2} \label{26}\\
			p_r^{eff}=\frac{(3+Cr^2)}{56\pi(1+Cr^2)^2}-D1 \label{27}\\
			p_t^{eff}=\frac{(3+Cr^2)}{56\pi(1+Cr^2)^2}-D1+\frac{15}{28}\frac{\beta Cr^2}{(1+Cr^2)^3} \label{28}
		\end{eqnarray}
		where $D1=\frac{4B}{3C}$ and $\beta=\frac{3}{4}$.
\section{Matching boundary conditions}\label{sec 5}
		As to ensure a well behaved stellar interior configurations for all compact structure which are mainly bounded objects, smooth geometry at the surface $r=R$ of the configuration need to be joined the inner space-time with the exterior space-time. The well-known Schwarzschild metric which is taken as exterior solution, can be written as \cite{Schwarzchild}
		\begin{equation*}
			ds^2=\big(1-\frac{2m}{r}\big)dt^2-\big(1-\frac{2m}{r}\big)^{-1}dr^2-r^2(d\theta^2+sin^2 \theta d\phi)
		\end{equation*}
		at the boundary $r=R$. With definition of continuity, the metric co-efficient $e^{\nu}$ and $e^{\lambda}$ across the surface boundaries $(r=R)$ and the radial pressure vanishes at the boundary, with equations, 
		\begin{equation*}
			1-\frac{2m}{r}=e^{\nu(R)}=e^{-\lambda(R)}
		\end{equation*}
		\begin{equation*}
			p_r(r=R)=0
		\end{equation*}
		On using the boundary conditions in the above equations, we obtained the value of arbitrary constants as
		\begin{eqnarray}
		    D1=\frac{(3+Cr^2)(\pi+3\xi)}{28\pi(8\pi+3\xi)(1+Cr^2)^2}-\frac{15\beta Cr^2\xi}{336\pi(8\pi+3\xi)(1+Cr^2)^3}\label{29}
		\end{eqnarray}
	    For solution of $Y$, 
	    \begin{eqnarray*}
	    	N1=\frac{log(Cr^2+1)}{6}-\frac{5log(4Cr^2+7)}{48}+\frac{7\pi(3log(4Cr^2+7)-4Cr^2)D1}{8}+\frac{3log(4Cr^2+7)}{16}
	    	\end{eqnarray*}
    	\begin{eqnarray*}
	    	CD=exp(N1)
	    	\end{eqnarray*}
    	\begin{eqnarray*}
	    	CN=\sqrt{\frac{7+4Cr^2}{7(1+Cr^2)}}
	    \end{eqnarray*}
    	\begin{eqnarray*}
	    	N2=\frac{CN}{CD}
	     \end{eqnarray*}
    \begin{eqnarray*}	
	    	N3=\frac{log(CR^2+1)}{6}-\frac{5log(4CR^2+7)}{48}+\frac{7\pi(3log(4CR^2+7)-4CR^2)D1}{8}+\frac{3log(4CR^2+7)}{16}
	     \end{eqnarray*}
    \begin{eqnarray*}	
	    	N4=exp(N3)
	    \end{eqnarray*}	
	    Therefore, 
	    \begin{eqnarray}
	    	Y=N4.N2\label{30}
	    	\end{eqnarray}

\section{Physical properties in $f(\Re,T)$ gravity}\label{sec 6}
		Here, all the physical requirements of the anisotropic stellar model were examined analytically as well as graphically to provide validation and stabilization of the system in the framework of $f(\Re,T)$ gravity theory.
		
\subsection{\bf{Metric functions, energy density and pressures}}\label{sec 6.1}
		In this paper, we have considered the metric potentials $e^{\lambda}$ and $e^{\nu}$ as in Buchdahl assumptions. In previous section \ref{sec 5}, the behavior of metric potential matched inner geometry to the exterior space-time in smoothly manner at $r=R$ to get value of constants. In regard to study the behavior of the main salient features of the model i.e density $\rho$, radial $p_r$ and tangential $p_t$ pressure respectively, the physical parameters of any compact object describing stellar structure should be monotonically decreasing towards the surface and should be maximum at the center. And the radial pressure should vanishes at the boundary whereas the energy density and tangential pressure are non-zero at the boundary. 	
		\par
\begin{figure}[h]
			\begin{center}
				\includegraphics[width=6cm]{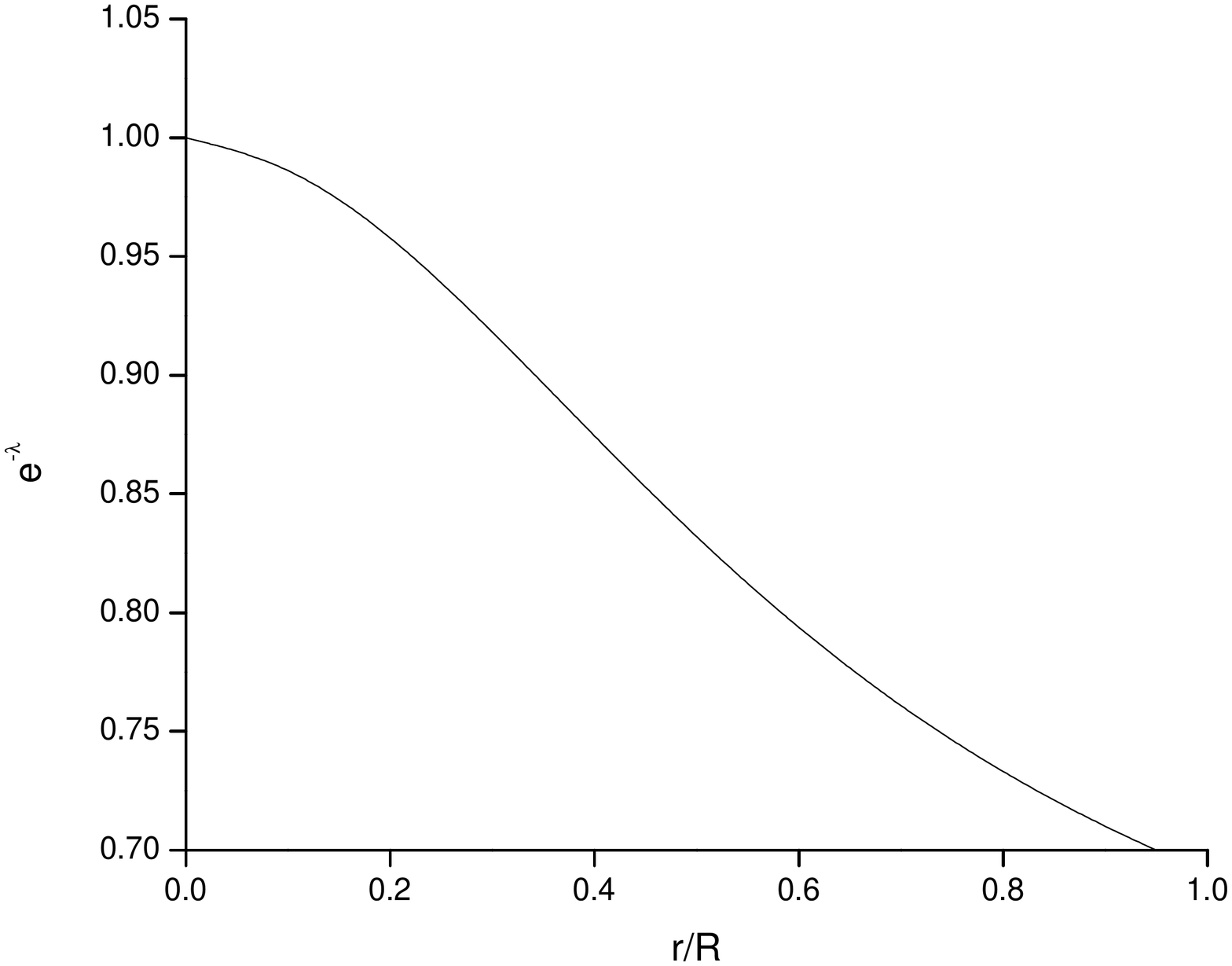}\includegraphics[width=6cm]{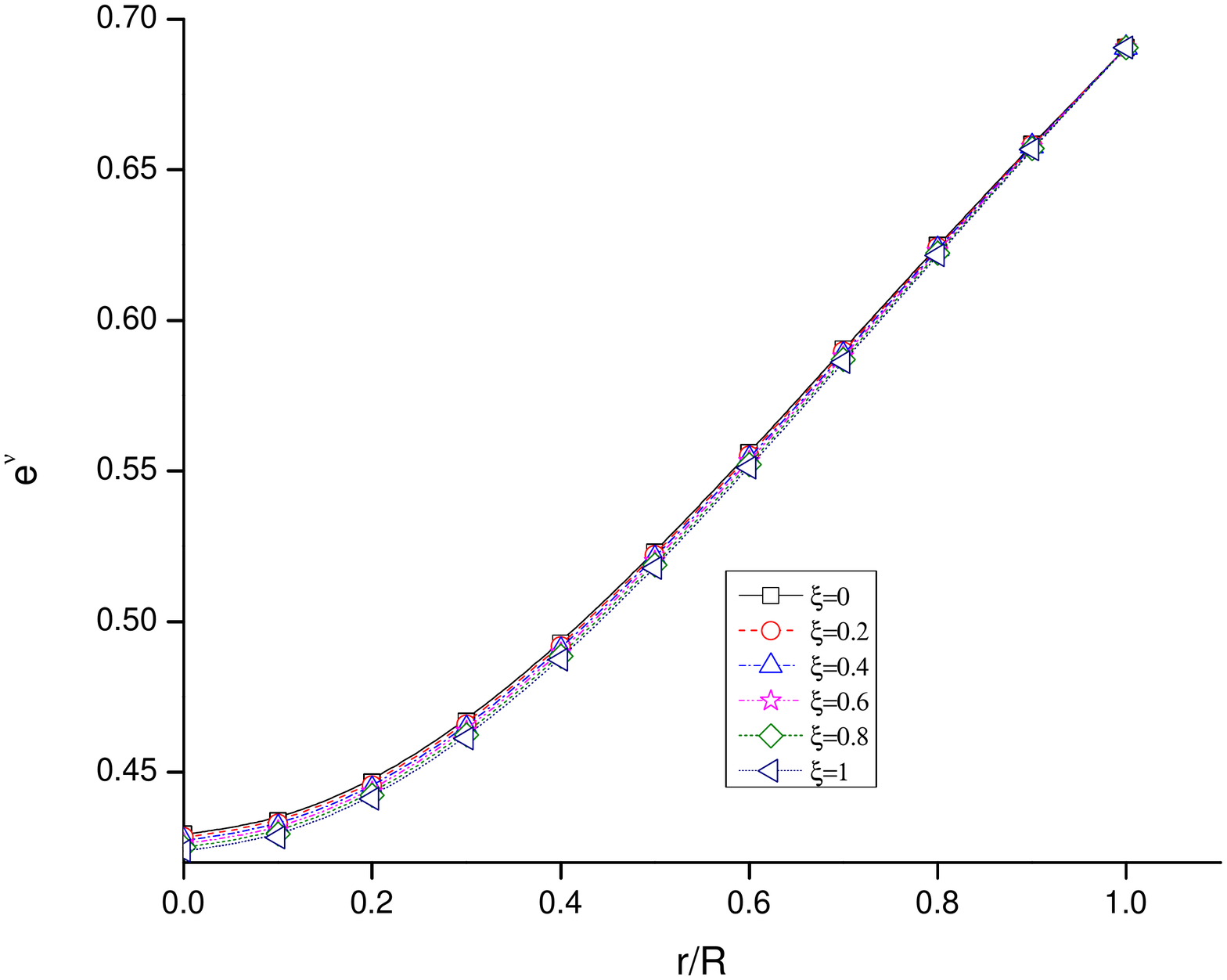}
				\caption{Above graphs represent metric potentials against radial coordinate ($r/R$) of compact star Her X-1 for different values of $\xi$.}\label{f1}
			\end{center}
		\end{figure}

\begin{figure}[h]
			\begin{center}
				\includegraphics[width=6cm]{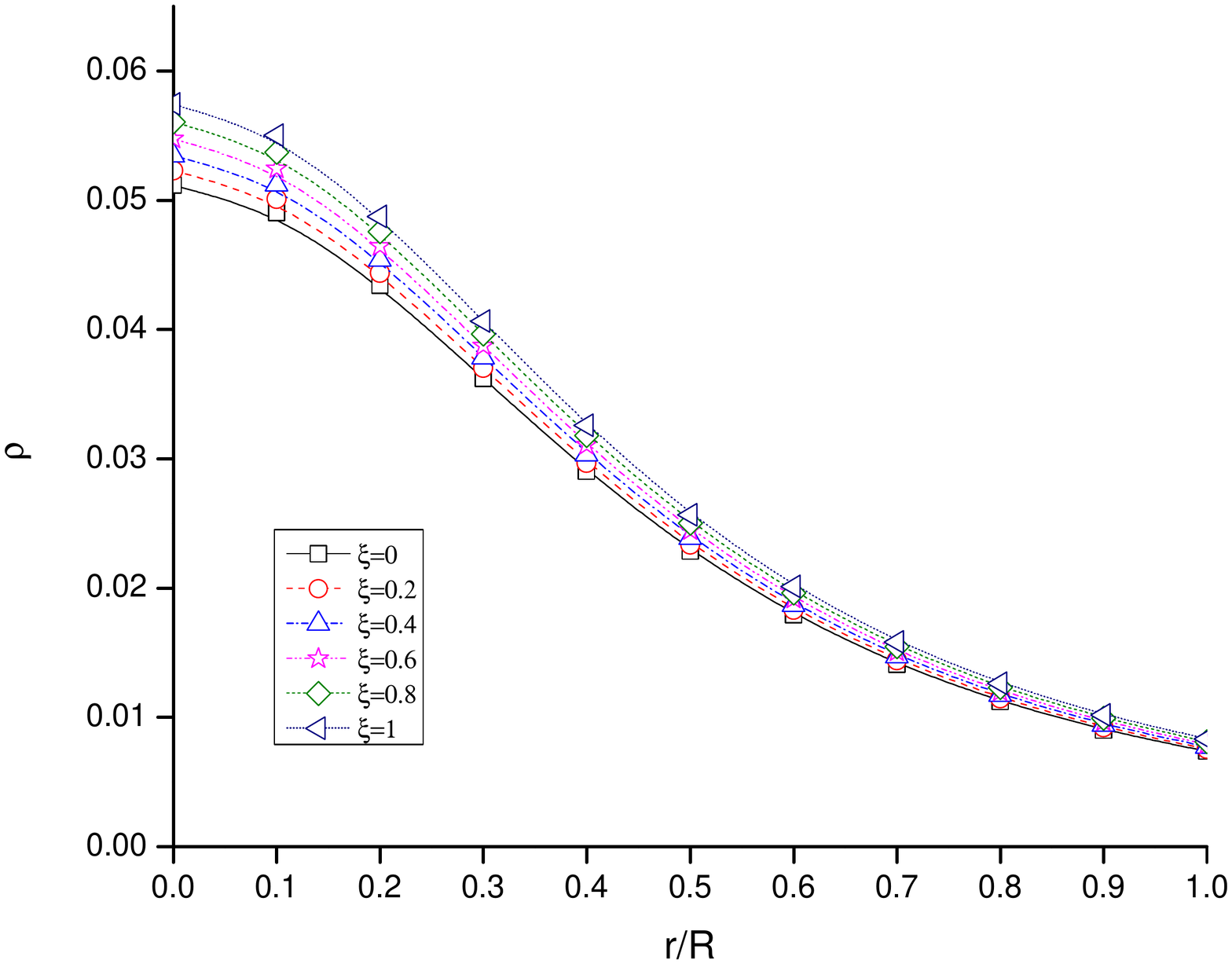}\includegraphics[width=6cm]{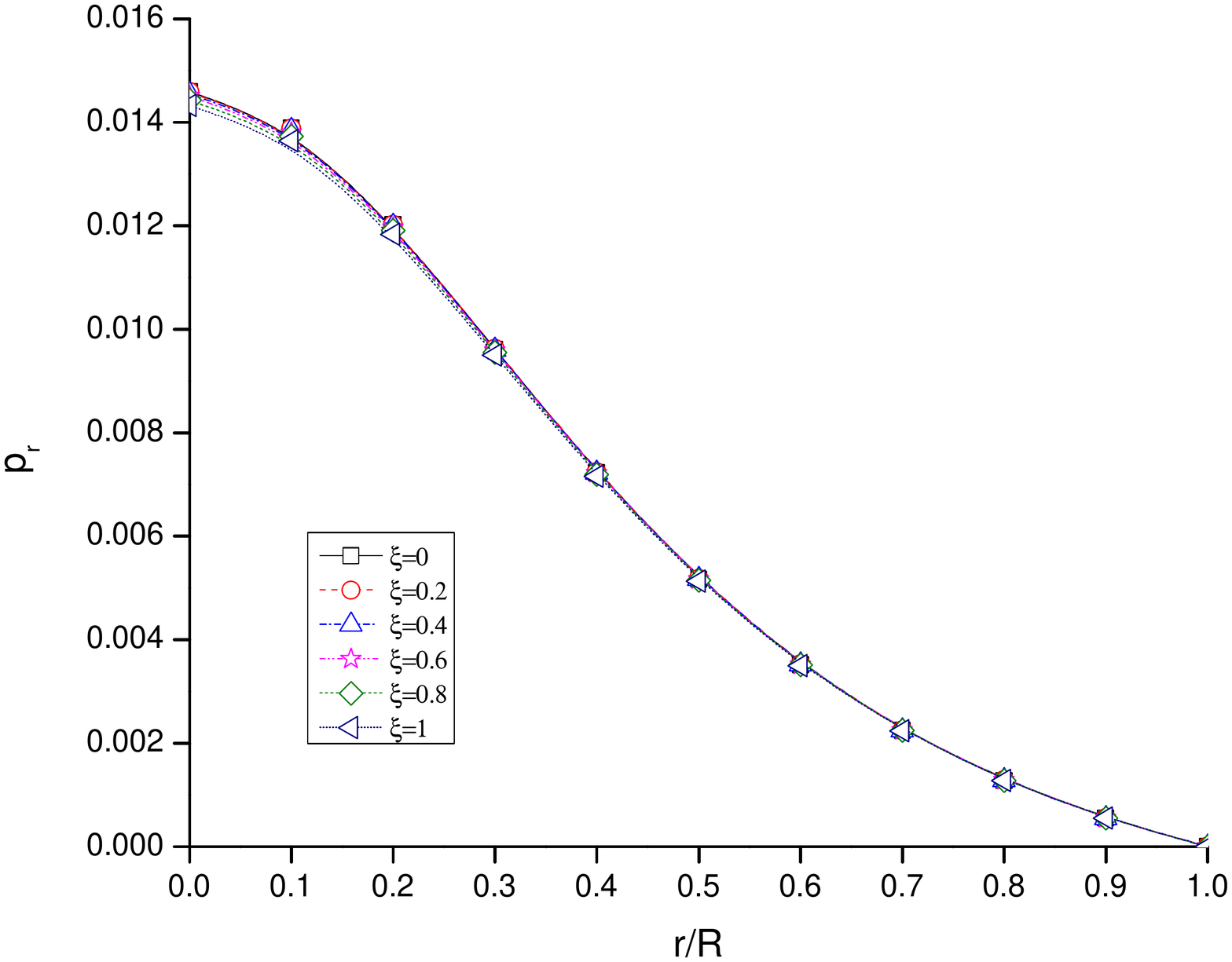}\includegraphics[width=6cm]{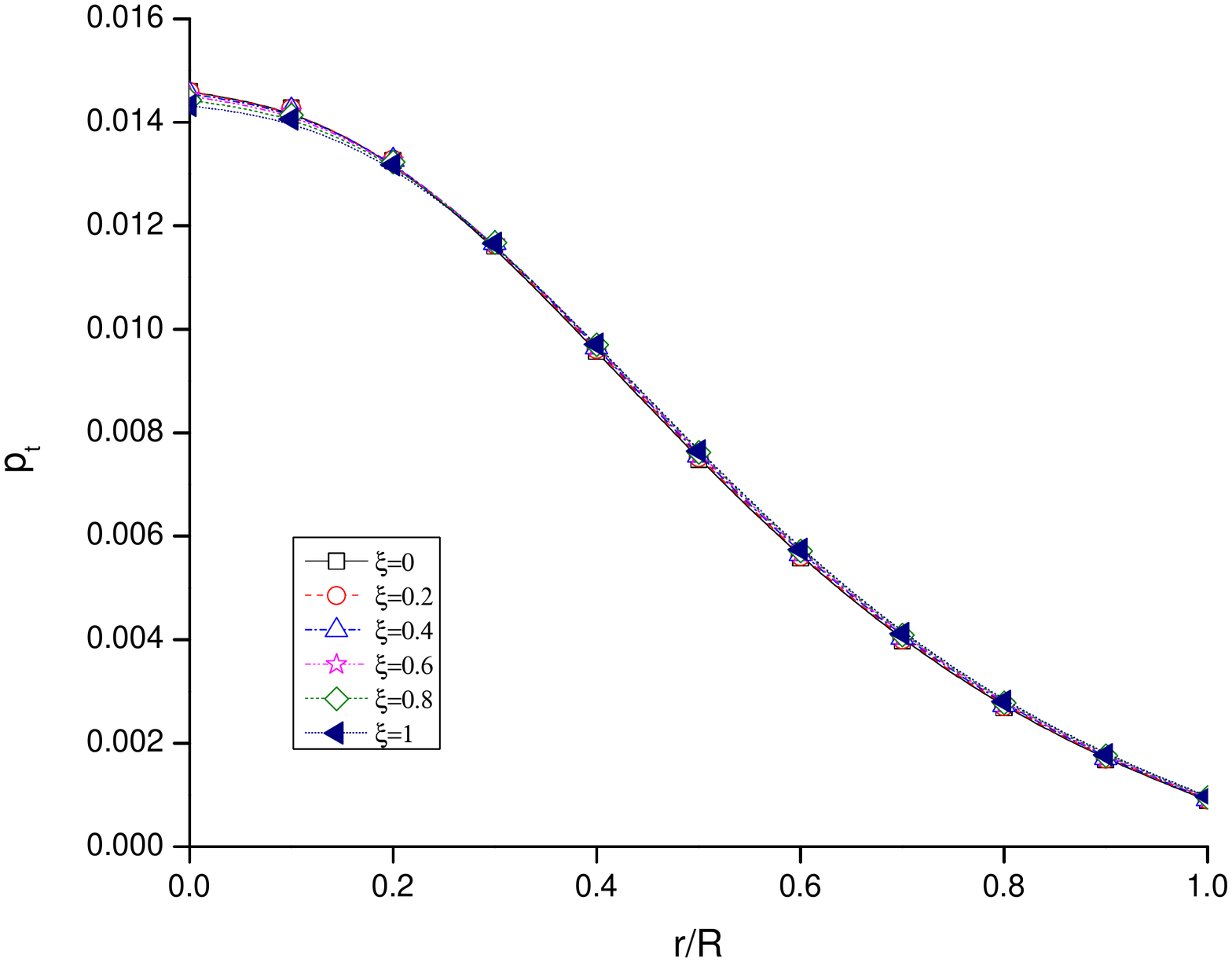}
				\caption{Above graphs represent energy density, radial pressure and tangential pressure respectively against radial coordinate ($r/R$) of compact star Her X-1 for different values of $\xi$.}\label{f2}
			\end{center}
		\end{figure}

 We have shown the graphical presentation of the metric functions in fig.\ref{f1}, density, and pressures ($p_r$ and $p_t$) in fig.\ref{f2} for different values of $\xi$.
\subsection{\bf{Anisotropy factor and ratio of radial and tangential pressure}}\label{sec 6.2}
In context of anisotrophy, the radial pressure is different from tangential pressure i.e, ${p_r}neq{p_t}$. So, the effective anisotropy factor is given as $\Delta=p_t-p_r$, this improves the stability and balancing mechanism and also increasing the value of the redshift. Depending on the signs of anisotropy, one can define the equilibrium mechanism contribution of it, i.e, where its positive, $\Delta>0 \implies p_t>p_r$  or negative, $\Delta<0 \implies p_t<p_r$. In first case, the system experiences a repulsive force which helps in counter balancing the gravitational gradient whereas in second case, the anisotropy force helps compress the object with gravitational force. And if the nuclear force is unable to overcome the gravitational force then the object forms a blackhole with unusual behaviors which means the attractive force because of anisotropy damages the balance and stability of the stellar model. It clearly confirms that the singularity depends on hydrostatic force exerted by the matter inside the star. The ratio of pressures i.e $\frac{p_r}{p_t}$ also confirms outward behavior of the anisotrophy factor.
\par
\begin{figure}[h]
			\begin{center}
				\includegraphics[width=6cm]{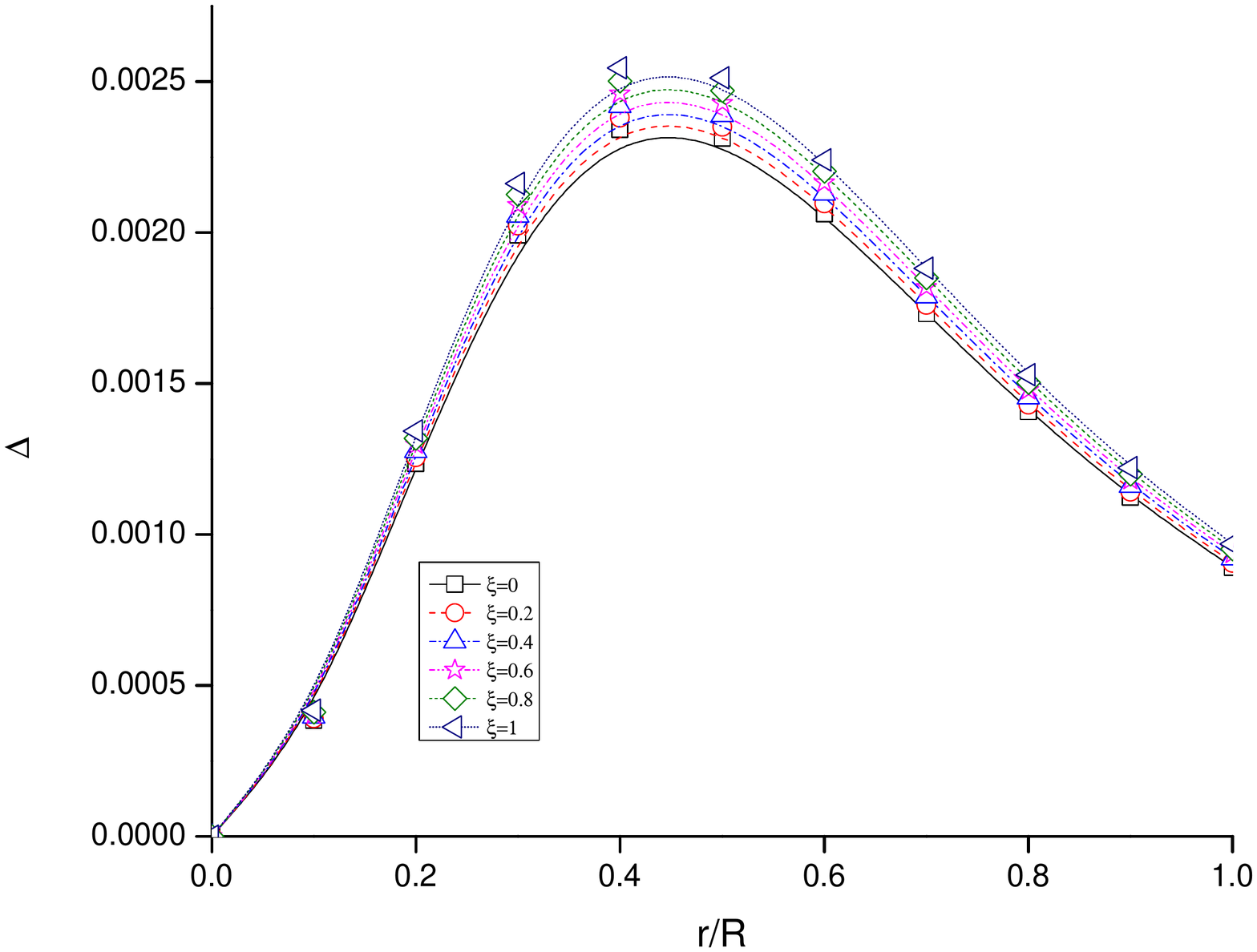}\includegraphics[width=6cm]{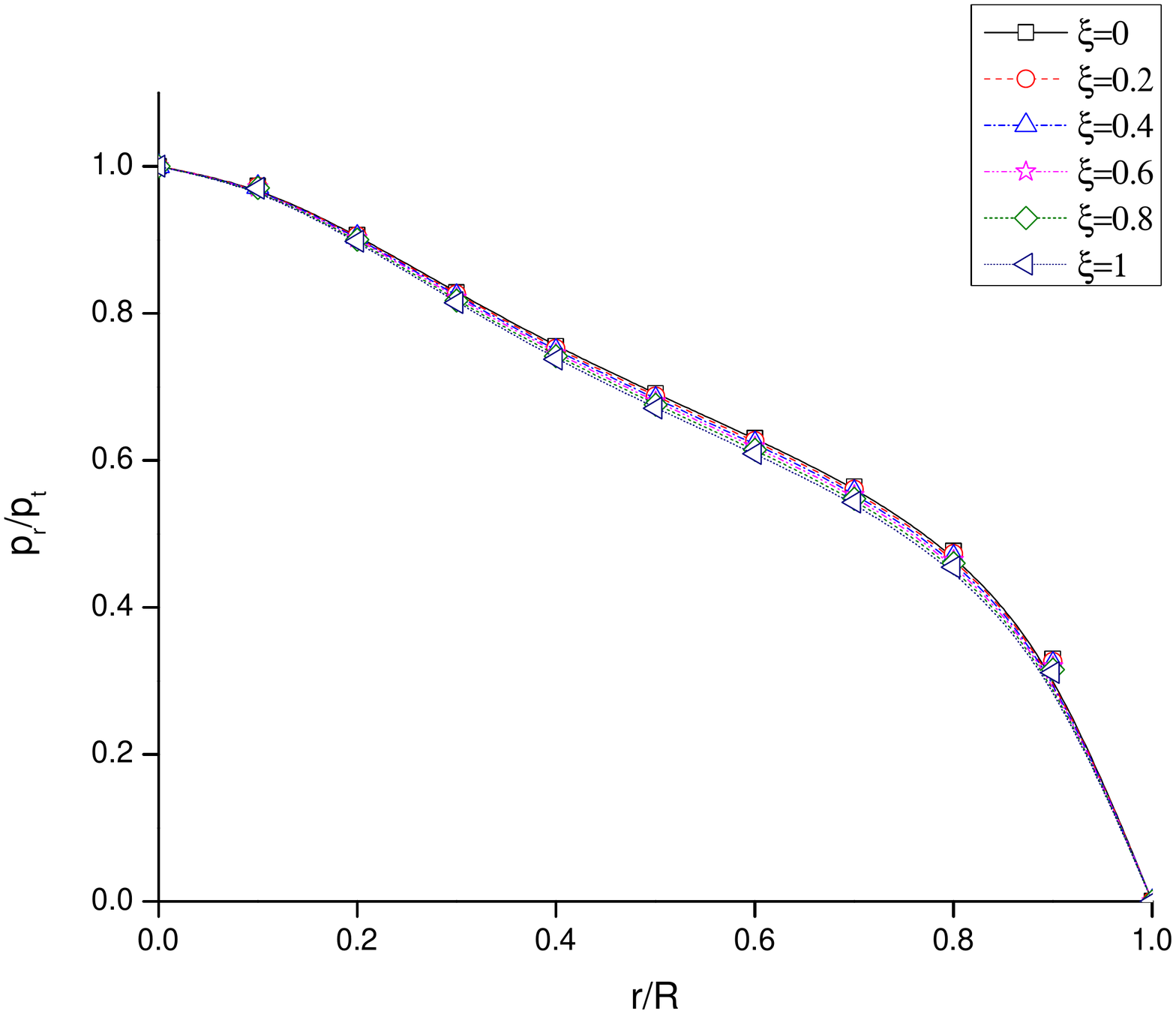}
				\caption{Above graphs represent anisotrophy factor and the ratio $\frac{p_r}{p_t}$ respectively against radial coordinate ($r/R$) of compact star Her X-1 for different values of $\xi$.}\label{f3}
			\end{center}
		\end{figure}

 We have graphically shown the behavior of anisotropy factor and ratio of pressures in fig.\ref{f3} for different values of $\xi$. 

\subsection{\bf{Energy conditions}}\label{sec 6.3}
\begin{figure}[h]
\includegraphics[width=5cm]{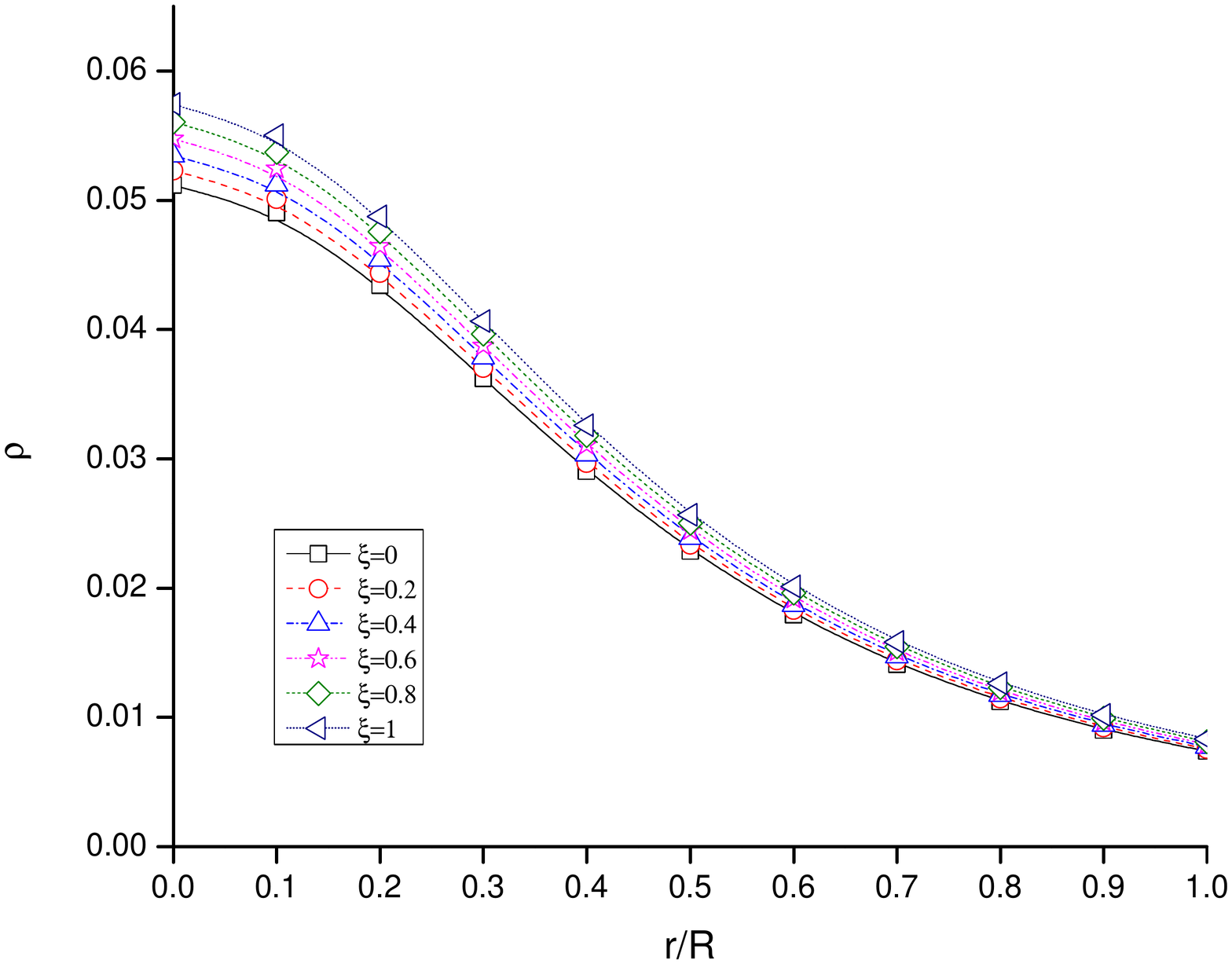}\includegraphics[width=5cm]{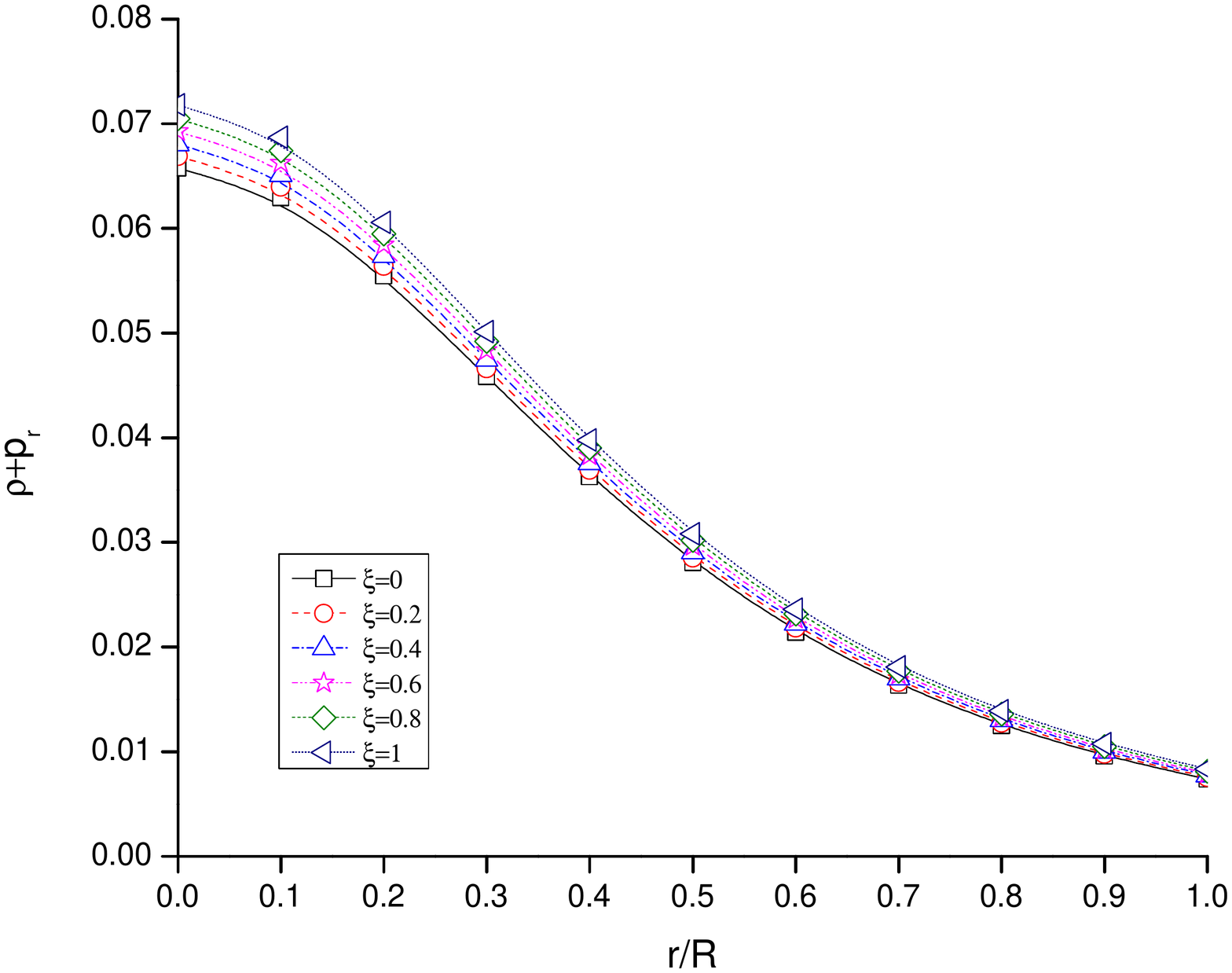}\includegraphics[width=5cm]{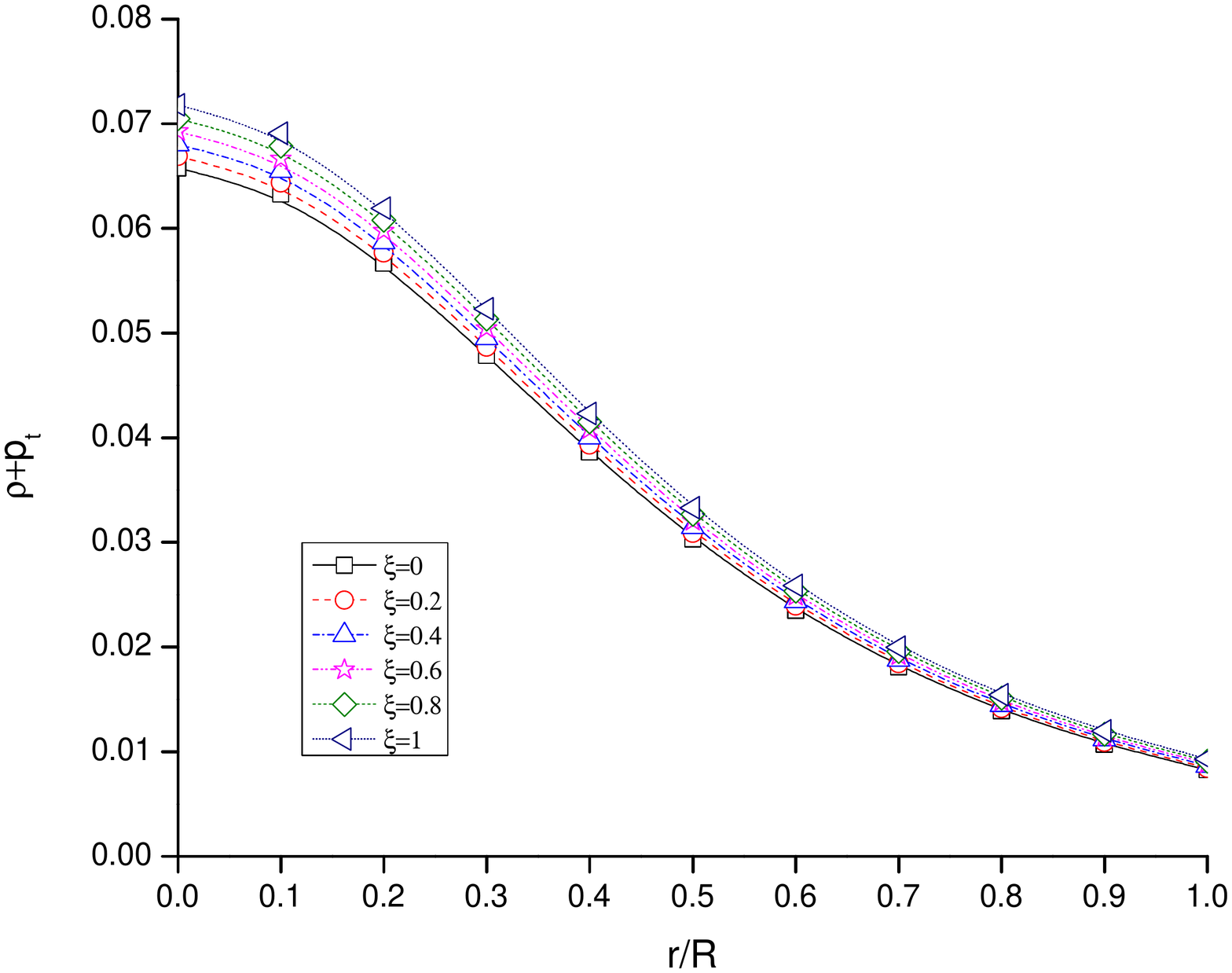}
\includegraphics[width=5cm]{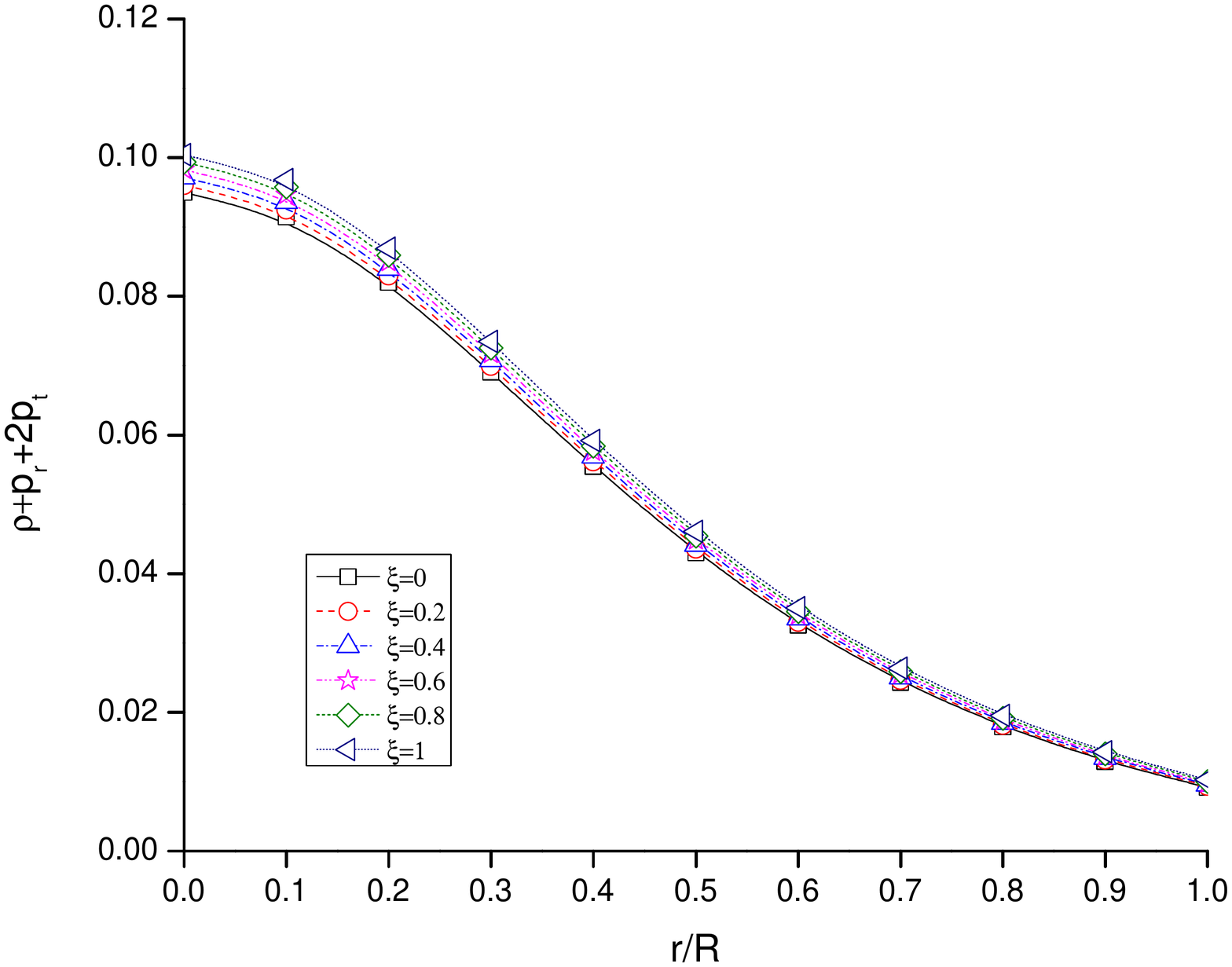}\includegraphics[width=5cm]{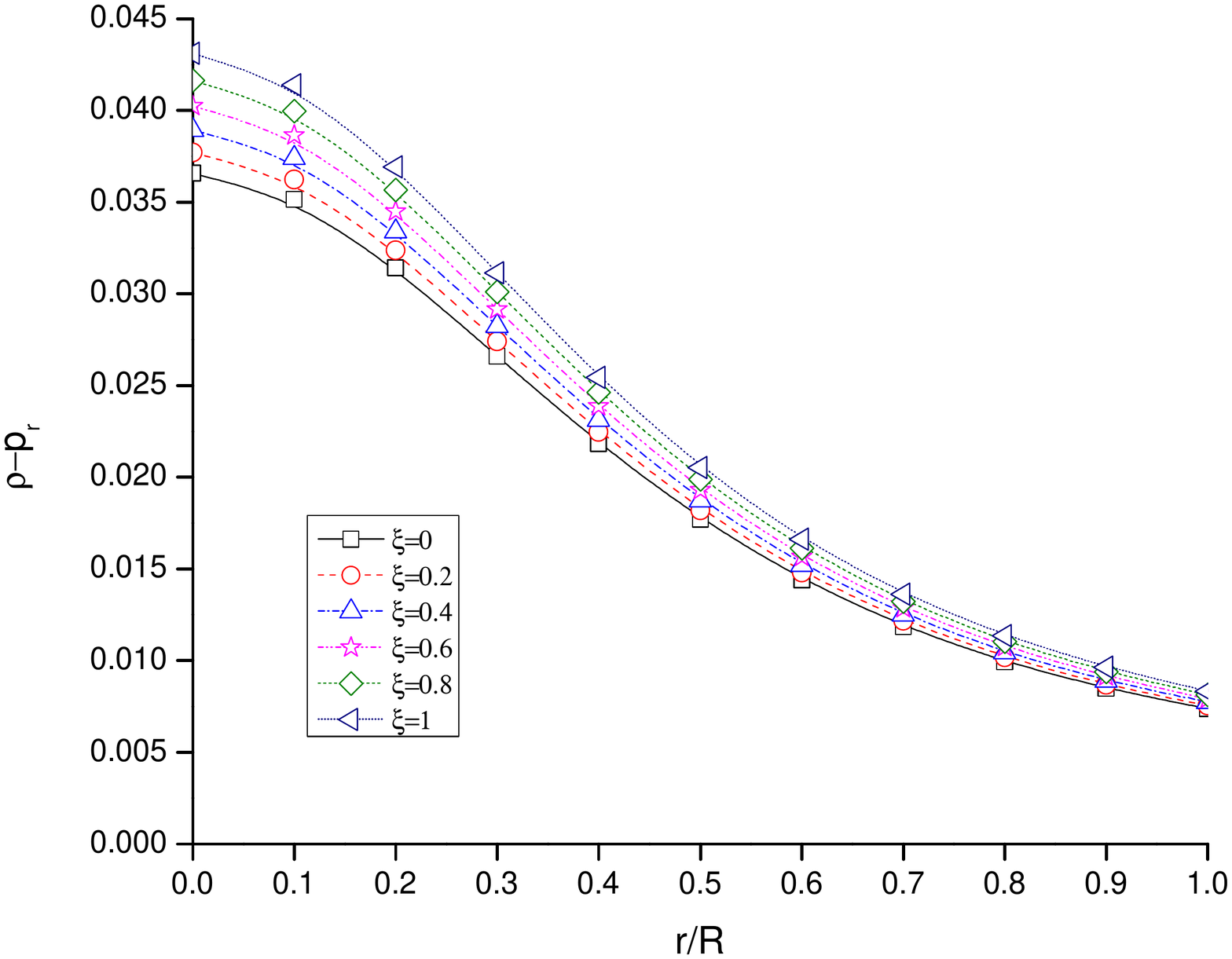}\includegraphics[width=5cm]{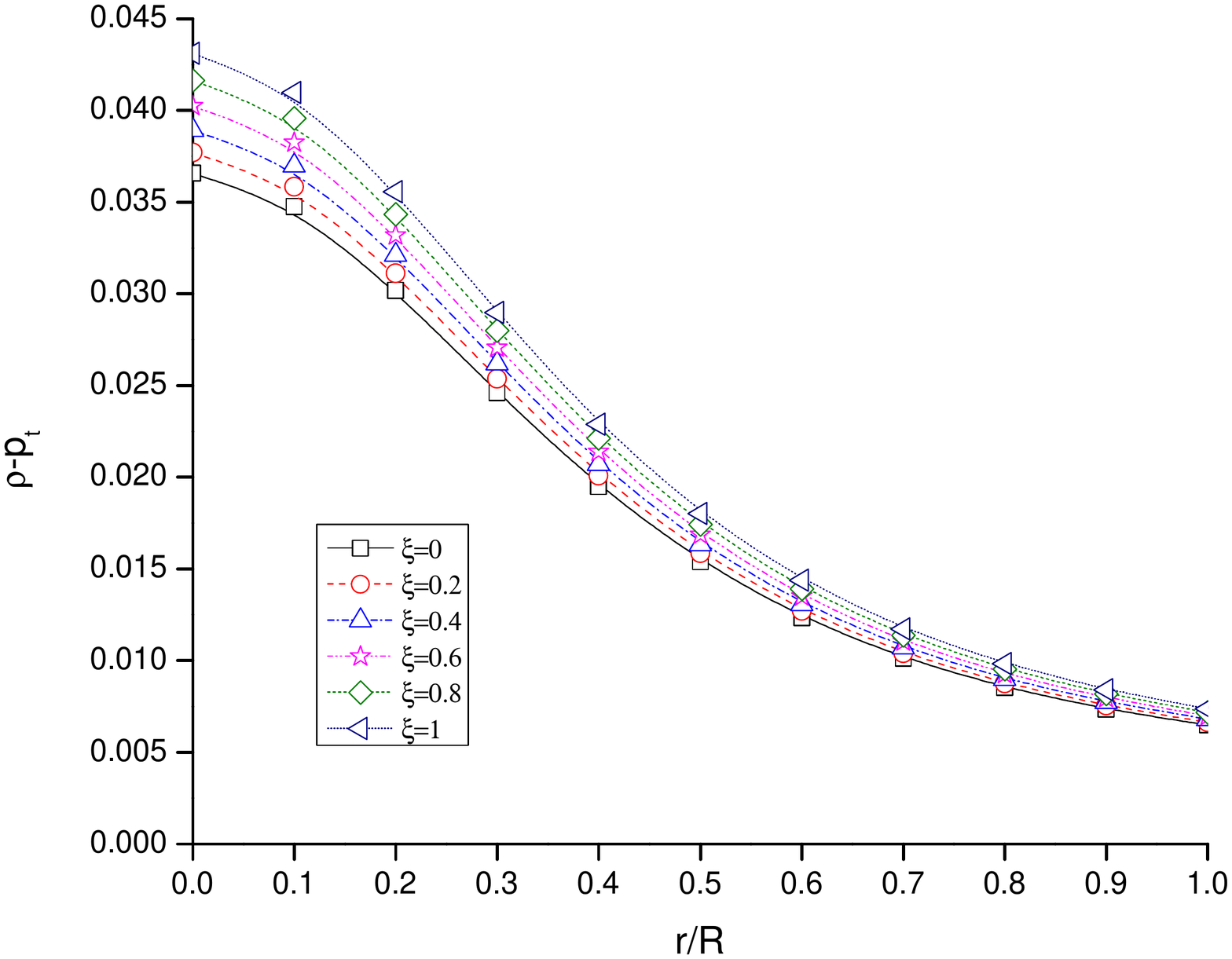}
\begin{center}
\includegraphics[width=5cm]{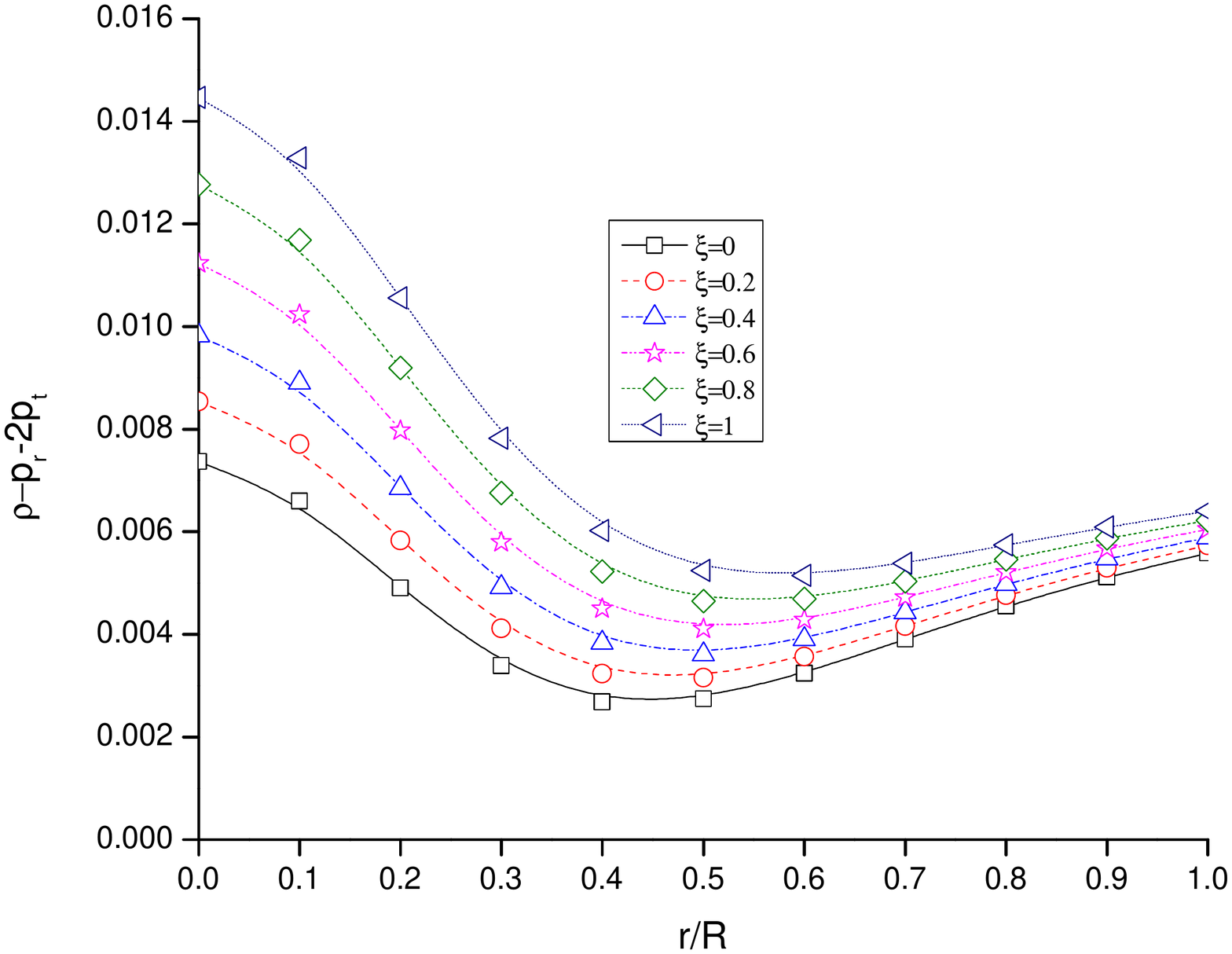}
\end{center}								
\caption{Above graphs represent different energy conditions  against radial coordinate ($r/R$) of compact star Her X-1 for different values of $\xi$}.\label{f4}
\end{figure}

		In this section, we verify the energy conditions for our choice $f(\Re,T)=\Re+2\xi T$ field equations. The authenticity of the energy condition is investigated by \cite{chakraborty} constraints namely, null energy condition (NEC), weak energy condition (WEC), strong energy condition(SEC), dominant energy condition(DEC) and lastly trace energy condition(TEC), which are mathematically given as
		\begin{align}
			& NEC: \rho(r) \geq 0 \nonumber\\
			& WEC: \rho(r)+p_i(r) \geq 0 \nonumber\\
			& SEC:\rho+p_r+2p_t\geq0 \nonumber\\
			& DEC: \rho(r)-p_i(r)\geq0 \nonumber\\
			& TEC: \rho(r)-p_r(r)-2p_t(r)\geq0 \nonumber
		\end{align}
		where $i$=radial $r$, transverse $t$. When an observer crosses the null diagram, NEC depicts that the quantity of  matter density is positive , when an observer transverses a time-like diagram, WEC indicates that the matter density is always positive whereas when tested by the relating observers, SEC depicts that for the observer the trace of the tidal tensor is always positive and DEC indicates that the mass-energy never flows faster than light and finally, TEC is basically the extension of SEC and DEC which indicates that the energy density is positive \cite{bekenstein}. Graphical behaviur of these energy conditions for different values of $\xi$ is provided in fig.\ref{f4}  that indicates the energy conditions are satisfying throughout the star surface for different values of $\xi$. 
		
\subsection{\bf{Compactness relation}}\label{sec 6.4}
\begin{figure}[h]
\begin{center}
\includegraphics[width=6cm]{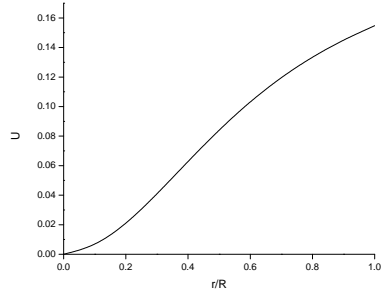}
\caption{Above graph represent compactness factor against radial coordinate of compact star Her X-1.}\label{f5}
\end{center}
\end{figure}

		For stellar structure, the value of mass function vanishes at the center of the star and is maximum at the boundary. In case of Buchdahl \cite{buchdahl}, the mass-radius ratio relation should be less than $\frac{8}{9}$. Mathematically, the mass function of the compact stellar structure is defined as 
		\begin{equation}
			m(r)=\frac{3Cr^3}{14(1+Cr^2)} \nonumber
		\end{equation}
		The compactness factor has an upper bound of $\frac{4}{9}$. For our present model compactness is obtained as 
		\begin{equation}
			u(r)=\frac{m(r)}{r}\label{31}
		\end{equation}
		Compactness factor are classified in different categories like if for normal star $u(r)=10^{-5}$ where as for white dwarfs it is $u(r)=10^{-3}$. In case of neutronstar, $u(r)\in (10^{-1}, \frac{1}{4})$, for ultra compact star $u(r)\in (\frac{1}{4}, \frac{1}{2})$ and for Black hole $u(r)=\frac{1}{2}$ \cite{rej}. Graphical representation of compactness is given in fig. \ref{f5} for different values of $\xi$.

\subsection{\bf{Gravitational and surface redshift}}\label{sec 6.5}
\begin{figure}[h]
\begin{center}
\includegraphics[width=6cm]{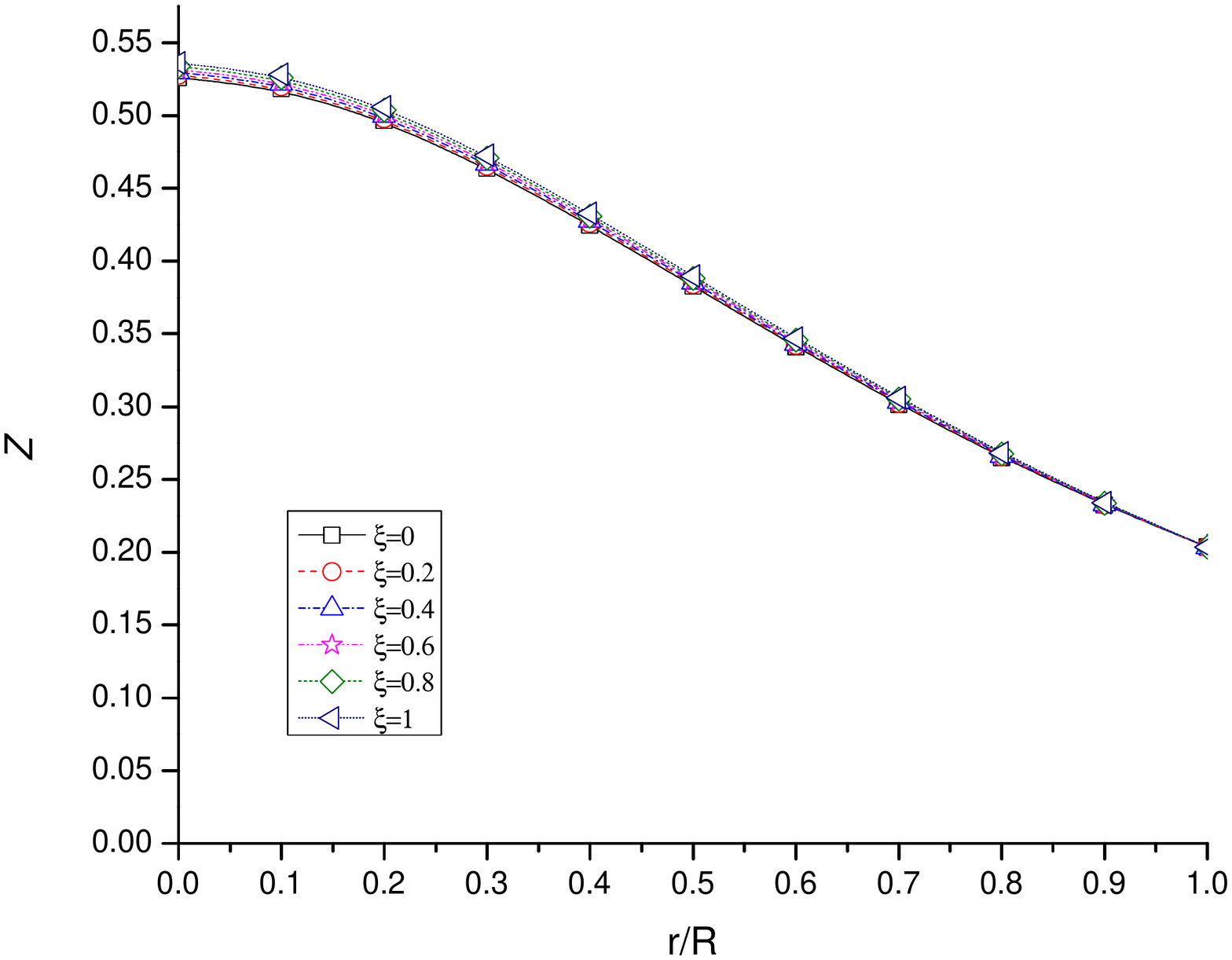}\includegraphics[width=6cm]{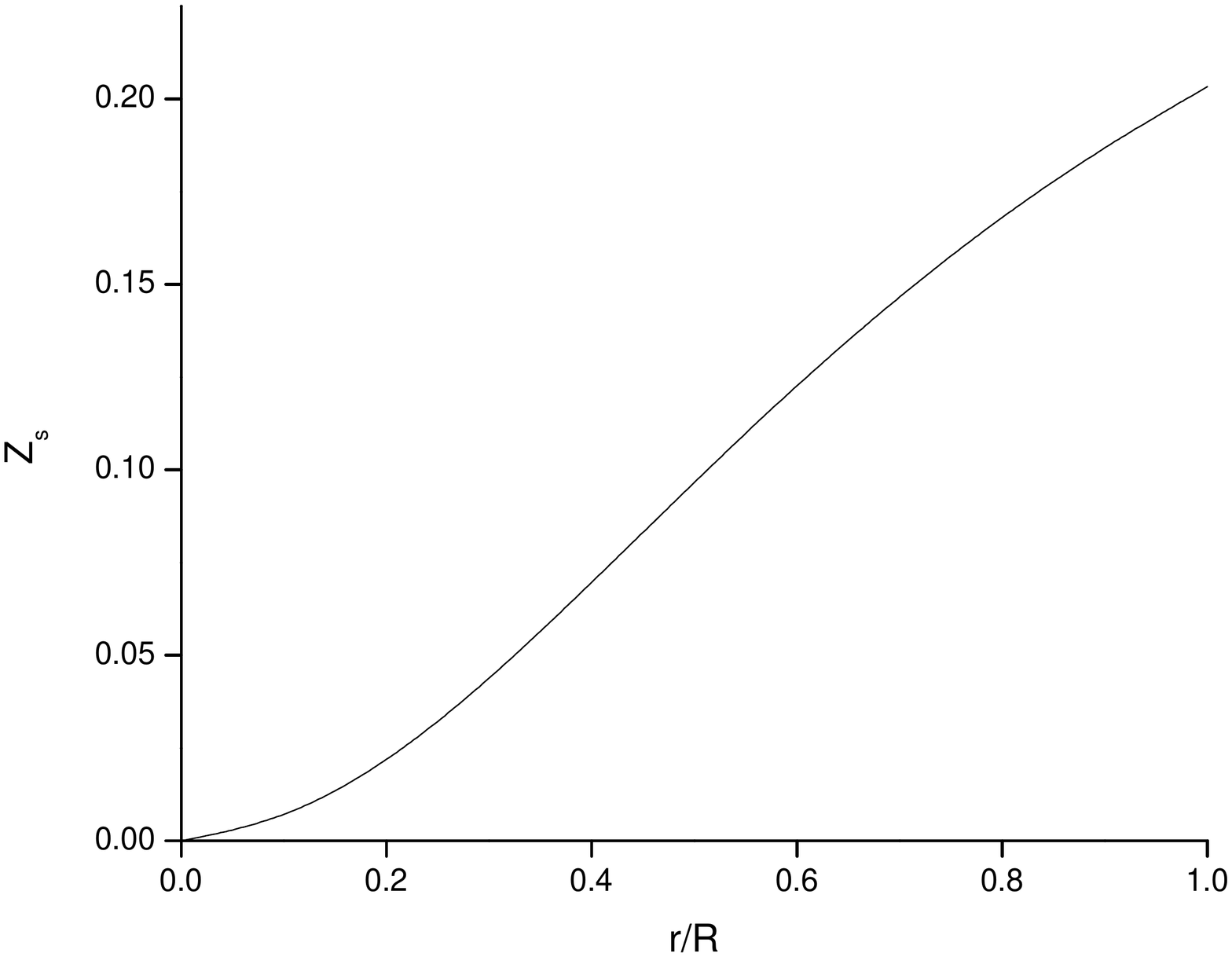}
\caption{Above graphs represent gravitational redshift  and surface redshift against radial coordinate of compact star Her X-1 for different values of $\xi$.}\label{f6}
\end{center}
\end{figure}

	 From definition of the gravitational redshift, $Z=\frac{\lambda_0-\lambda_e}{\lambda_e}$ where $\lambda_0$ is the observed wavelength at a distance $r$ and $\lambda_e$ is the emitted wavelength from the surface of the compact object \cite{deb}. So, the redshift is defined as
		\begin{equation}
			Z=\sqrt{e^{-\nu(R)}}-1 \label{32}
		\end{equation}
The surface redshift depends on the mass and radius of stellar object. With increase in mass, radius increases which increases surface gravity.
Thus, the surface redshift is given by
\begin{equation}
			Z_s=\frac{1}{\sqrt{1-2u}}-1 \label{33}
		\end{equation}
 Graphical representation of the gravitational redshift is given in fig. \ref{f6} for different values of $\xi$.
		
\subsection{\bf{Causality condition}}\label{sec 6.6}
\begin{figure}[h]
\begin{center}
\includegraphics[width=6cm]{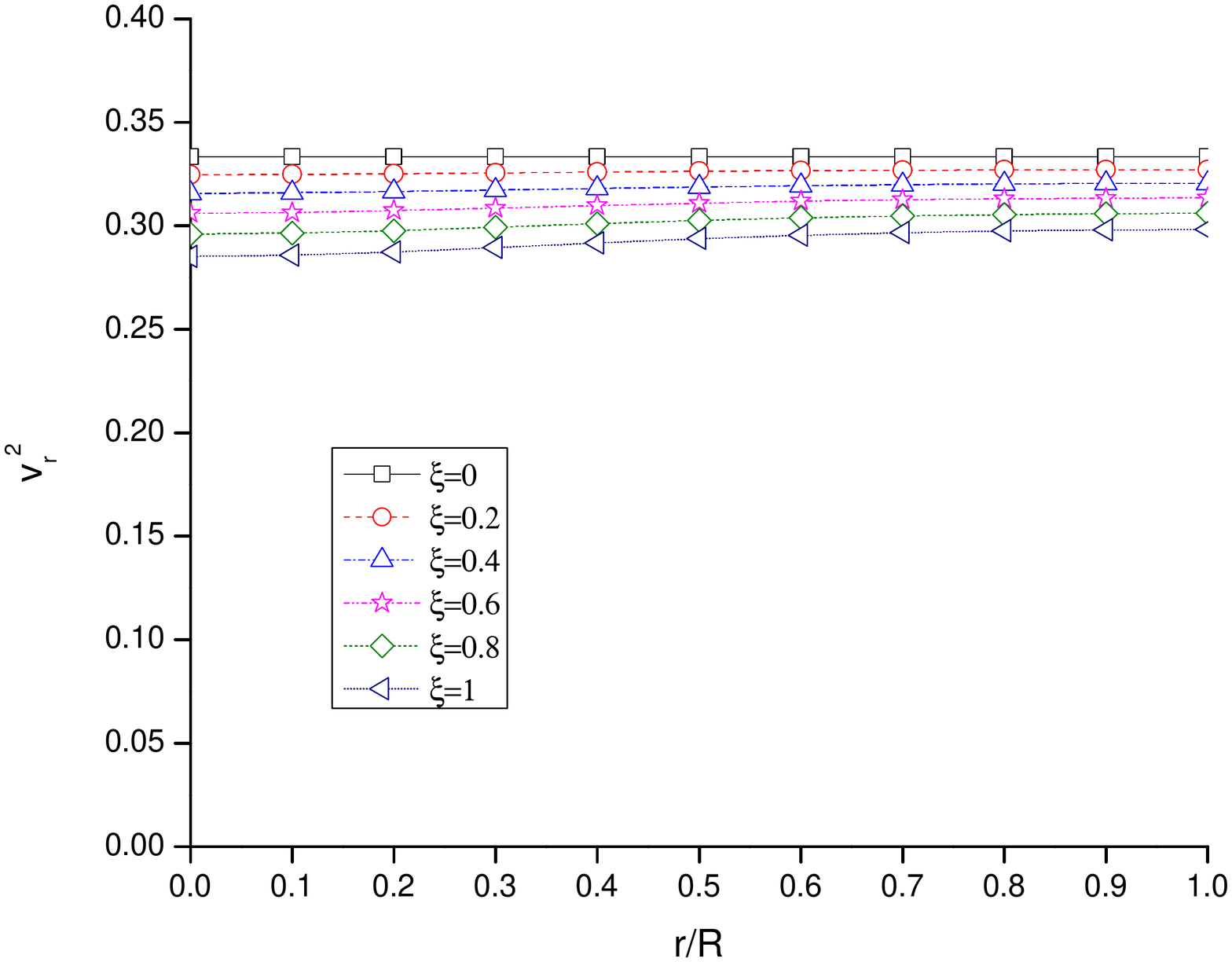}\includegraphics[width=6cm]{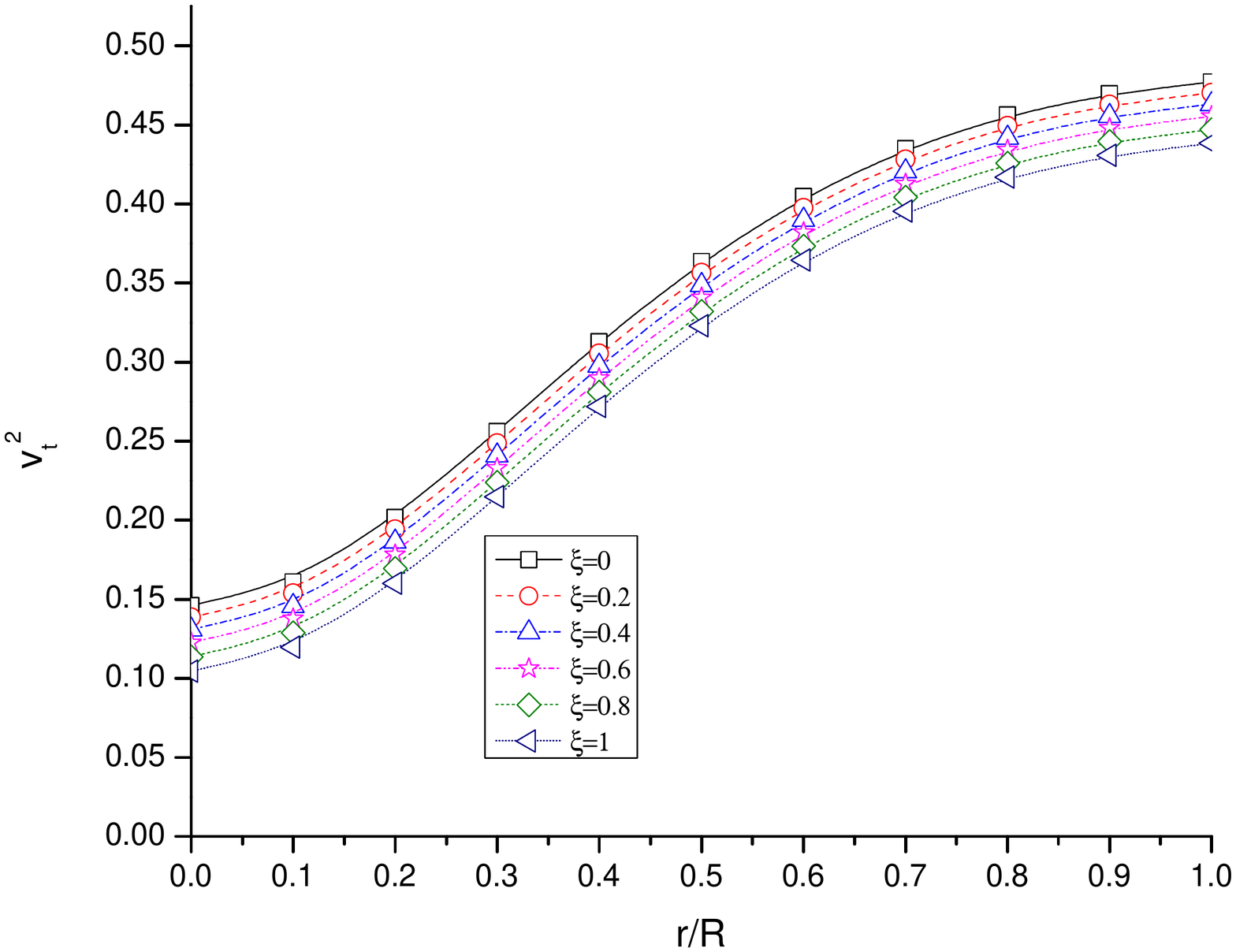}
\caption{Above graph represent causality condition against radial coordinate of compact star Her X-1 for different values of $\xi$.}\label{f7}
\end{center}
\end{figure}

		According to Herrera \cite{herrera}, physical acceptability of a model can be studied from the speed of sound. For stable stellar structure, square of radial $(v^2_r)$ and tangential $(v^2_t)$ should be less than 1 \cite{abreu}. It basically tells that the sound waves do not travel at arbitrary speeds and this know as causality condition. Regardless of the material content of the star the only difference is that in case of anisotropic, the propagation is in two directions of the sphere i.e, radial and transverse directions where as in case of isotropic the speed of sound decreases subliminally. So, the causality condition given by $0\leq v_r \leq 1$ and $0\leq v_t \leq 1$. Mathematically, these velocities define as 
		\begin{align}
			v^2_r=\frac{dp_r}{d\rho} \label{34}\\
			v^2_t=\frac{dp_t}{d\rho}\label{35}
		\end{align}  
		For $\xi>0$, the radial propagation is greater than tangential propagation. Similarly, for $\xi<0$, both the propagation are decreasing in nature. Graphically, we represented  both causality conditions in fig. \ref{f7}, which confirms the stablity of the system for all $\xi$.
		
\subsection{\bf{Herrera-cracking concept}}\label{sec 6.7}
\begin{figure}[h]
\begin{center}
\includegraphics[width=6cm]{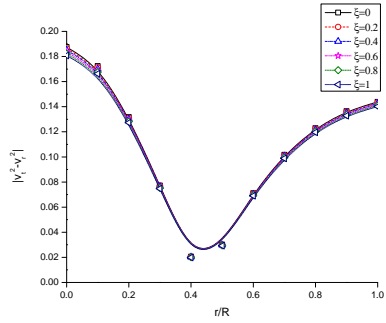}
\caption{Above graph represent stability factor against radial coordinate of compact star Her X-1 for different values of $\xi$.}\label{f8}
\end{center}
\end{figure}

		Another method for checking the stability condition is by Herrera-Cracking condition, which basically determines the stability of the system by taking variation in distribution of speed of sound into account. The differences between both velocities should be less than 1 i.e 
		\begin{equation*}
			|v^2_t-v^2_r|\leq 1
		\end{equation*}
		The graphical representation in fig. \ref{f8} shows that the obtained stellar system satisfies the stability condition.
		
\subsection{\bf{Adiabatic index}}\label{sec 6.8}
\begin{figure}[h]
\begin{center}
\includegraphics[width=6cm]{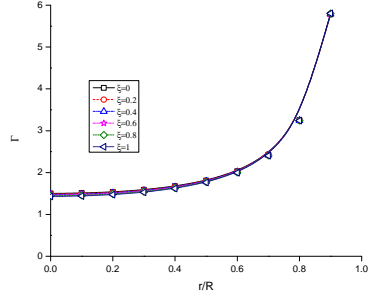}
\caption{Above graph represent adiabatic index against radial coordinate of compact star Her X-1 for different values of $\xi$.}\label{f9}
\end{center}
\end{figure}

		For a relativistic anisotropic, the stellar configuration of a compact body with spherically symmetric configuration is examined by the adiabatic index as it defines the intensity of equation of state at a defined density \cite{harrison}\cite{haensel}. Many authors \cite{hillebrandt}\cite{horvatz}\cite{doneva} provided different methods to check the stability of the compact objects by following the work of Chandrasekhar \cite{chandrasekhar}. In work of Heintzmann and Hillebrandt \cite{heintzmann} for the proposed spherically symmetric model the stability of the configuration holds when $\Gamma> 4/3$, where $\Gamma$ is the adiabatic index given as
		\begin{align}
			\Gamma = \frac{\rho+p_r}{p_r}\frac{dp_r}{d\rho} \nonumber\\
				= \frac{\rho+p_r}{p_r}v^2_r\label{36} 
		\end{align} 
		The graphical presentation of adiabatic index is given in fig. \ref{f9} which shows that the stability of $\Gamma$ and is greater than $4/3$ for all values of $\xi$.
		
\subsection{\bf{EOS parameter}}\label{sec 6.9}
\begin{figure}[h]
\begin{center}
\includegraphics[width=6cm]{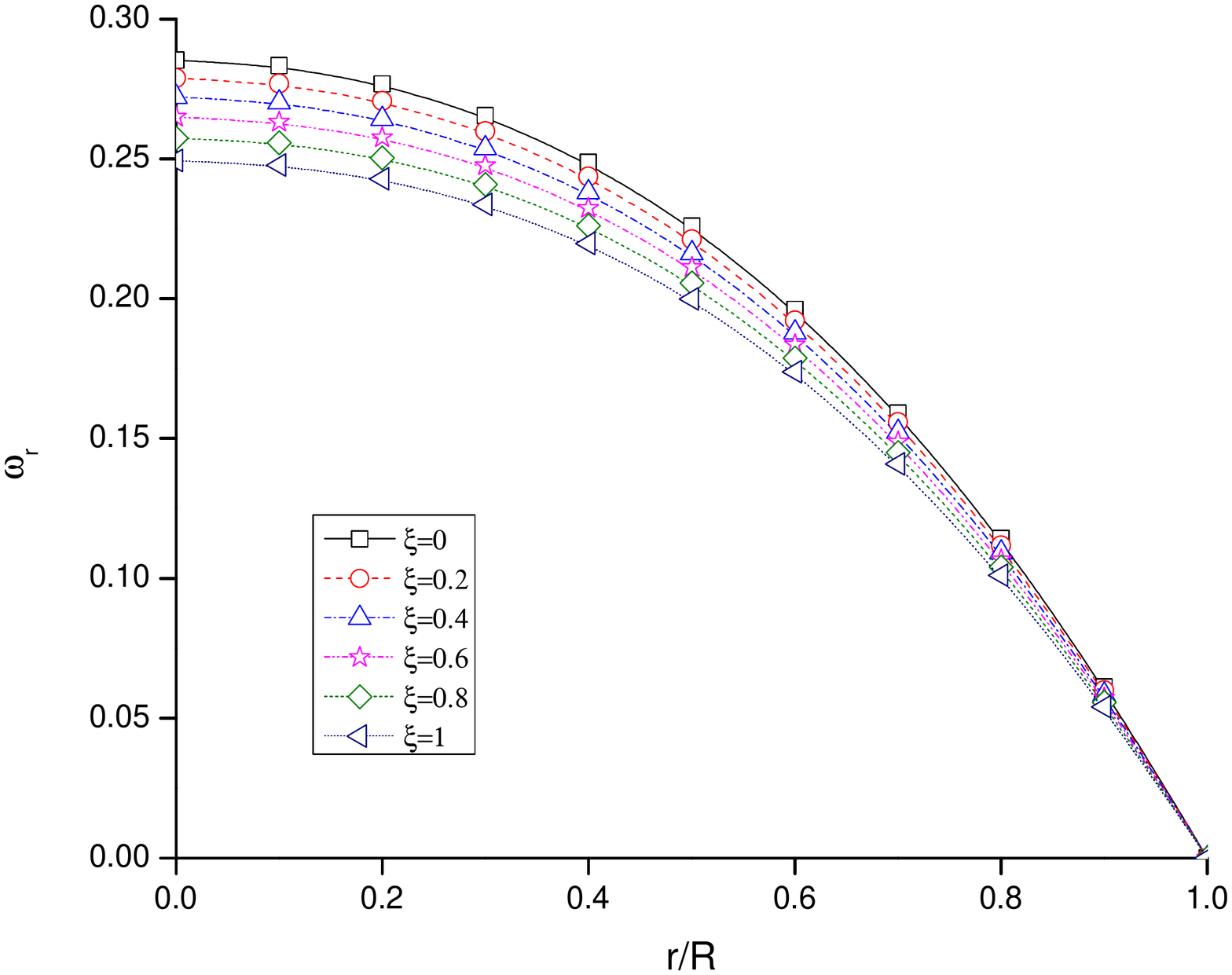}\includegraphics[width=6cm]{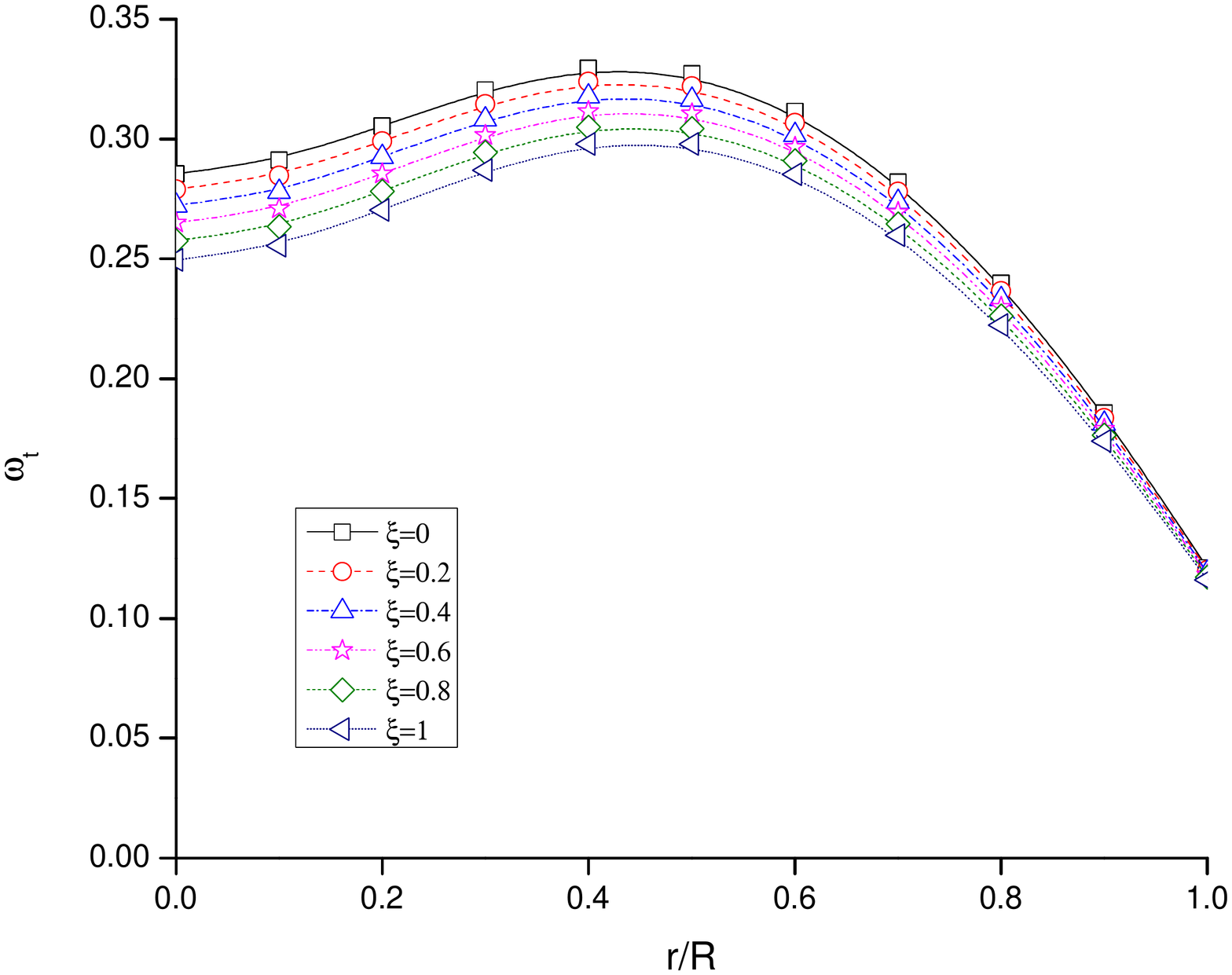}
\caption{Above graphs represent equation of state(EOS) parameters against radial coordinate of compact star Her X-1 for different values of $\xi$.}\label{f10}
\end{center}
\end{figure}

		Here, we define and check the nature of radial and transverse equation of state parameters. Expressions defining  equation of state parameters are given as follows
			\begin{align}
				W_r=\frac{p_r}{\rho}\label{37} \\
				W_t=\frac{p_t}{\rho}\label{38}
			\end{align}
		where $p_r$ and $p_t$ indicates the radial and tangential pressures whereas $\rho$ indicates the density of the anisotropic fluid. The graphical presentation of the EOS parameters is given in fig. \ref{f10}, which showed the behaviour of both parameters to be monotonically decreasing for different values of $\xi$.

\subsection{\bf{TOV equation}}\label{sec 6.10}
\begin{figure}[h]
			\begin{center}
				\includegraphics[width=6cm]{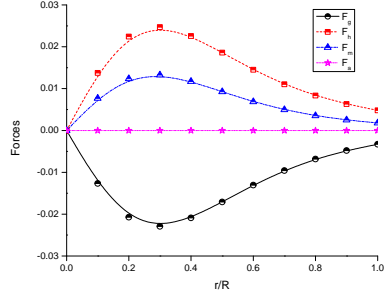}
				\caption{Above graphs represent the behavior of gravitational force $F_g$, hydrostatic force $F_h$, anisotropic force $F_a$ and force due to modified theory $F_m$ against radial coordinate of compact star Her X-1 for different values of $\xi$.}\label{f11}
			\end{center}
		\end{figure}

		In this section, we have undergone checking the equilibrium condition of the stellar model by Tolman-Oppenheimmer-Volkoff(TOV) \cite{tolman}\cite{volkoff}, using different forces acting on our present system in the framework of $f(\Re,T)$ theory gravity. As we know that the Einstein field equations for $f(\Re,T)$ gravity theory which leads to the energy conservation for our stellar model as $\bigtriangledown^{\mu}T_{\mu\nu}=0$. So, the conservation equation of the energy-momentum tensor which implies the modified form of the TOV equation in context of $f(\Re,T)$ gravity theory for our system as follows
		\begin{equation}
				-\frac{dp_r}{dr}-\frac{1}{2}\nu^{\prime}(\rho+p_r)+\frac{2}{r}(p_t-p_r)+\frac{\chi(3\rho^{\prime}+p_{r}^{\prime}+2p_{t}^{\prime})}{3(8\pi+2\chi)} =0\label{39}
		\end{equation}
		Where as the equilibrium equation has four forces namely: the hydrostatic force $F_h$, the gravitational force $F_g$, the anisotropic force $F_a$ and finally the force related to modified gravity $F_m$. However, the explicit form of these forces are given by
		\begin{align}
			& F_h = \frac{-dp_r}{dr}\label{40}\\
			& F_g = \frac{\nu^{\prime}}{r}(\rho+p_r)\label{41}\\
			& F_a = \frac{2}{r}(p_t-p_r)\label{42}\\
			& F_m = \frac{\chi(3\rho^{\prime}+p_{r}^{\prime}+2p_{t}^{\prime})}{3(8\pi+2\chi)}\label{43}
		\end{align}
		For $\xi=0$, the conservation equation in Einstein-Maxwell gravity. The modified TOV equation describes that the sum of the different forces in our stellar model should be zero i.e
		\begin{equation}
			F_g+F_h+F_a+F_m=0\label{45}
		\end{equation}
The behaviour of these forces is graphically represented in fig. \ref{f11} for different values of $\xi$. 		

\begin{figure}[h]
			\begin{center}
				\includegraphics[width=6cm]{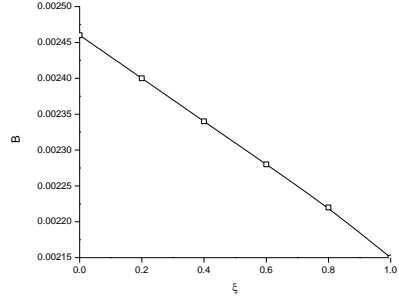}
				\caption{Above graphs represent the behavior of bag constant  against different values of $\xi$ of compact star Her X-1.}\label{f12}
			\end{center}
		\end{figure}
\begin{table}
	\caption{Values of the model parameters of central density, surface density, central radial pressure, adiabatic index, surface redshift, compactness and bag constant for different values of $\xi$ for the compact star Her X-1 by assuming $M=0.85M_{\odot}$ and $R=8.1km$. }\label{T1}
	\begin{center}
	\begin{tabular}{cccccccc}
		\toprule
		  $ \xi $ & $  \rho_{c} $ & $ \rho_{s} $ & $p_{r_c}$ & $ \Gamma_{r=0} $& $Z_s(R)$ & $U(R)$ & $B$\\
		\midrule
	      $ 0 $ & $ 0.051157 $ & $ 0.00737 $ & $ 0.014596 $ & $ 1.5106154 $ & $0.203423$& $0.154750$& $ 0.002457$ \\
		
		  $ 0.2 $ & $ 0.052289 $ & $ 0.007546 $ & $ 0.014583 $ & $ 1.489027 $ & $0.219592$ & $0.164218$ & $ 0.002402 $  \\
		
	      $ 0.4 $ & $ 0.053479 $ & $ 0.00773 $ & $ 0.014553 $ & $ 1.475531 $ & $0.235761$& $0.173686$ & $ 0.002342 $ \\
		
		  $ 0.6 $ & $ 0.054731 $ & $ 0.007923$ & $ 0.014502 $ & $ 1.461064 $ & $0.251930$& $0.183154$& $ 0.002282 $ \\
		
	      $ 0.8 $ & $ 0.05605 $ & $ 0.008126 $ & $ 0.014428 $ & $ 1.445505 $ & $0.268099$& $0.192622$& $ 0.002218 $ \\
		
		  $ 1 $ & $ 0.057444 $ & $ 0.00834 $ & $ 0.014326 $ & $ 1.428709 $ & $0.284268$& $0.202090$& $ 0.002151 $ \\
		\bottomrule
	\end{tabular}
\end{center}	
\end{table}

\section{Discussion and conclusion}\label{sec 7} 
		In this exposition, we have investigated a new class of general solutions for the spherically symmetric anisotropic stellar structure. To our best knowledge, the investigation is based on the simplified form of $f(\Re,T)$ function. We have employed the physically motivated Buchdahl ansatz\cite{buchdahl} for the metric potentials. Our system is governed by simplified phenomenological MIT bag model equation of state, i.e $p_r=\frac{1}{3}(\rho-4B)$. We have choose $f(\Re,T)=\Re+2\xi T$ basing on it we put our model to rigorous tests for checking its regularity, causality and stability which emphasizes on the role of $\xi$, coupling constant. It has been shown that our system conserves energy-momentum tensor in the framework of $f(\Re,T)$ gravity theory. Through the boundary conditions, we fixed the free constants which are arises due to integration of the field equations. In connection to features, we bring out the contributions from the modified theory we have explored several physical aspects based on our findings for various values of $\xi$, coupling constant. All these have related very interesting advocacy in favour of physical acceptance of the model. We have illustrated the observations graphically for Her X-1 as the representative of he compact stars.
		
		\par Here, we have presented characteristics of the metric potentials by $e^{\nu}$ and $e^{\lambda}$ in fig.\ref{f1}, which depicts that our stellar configuration is free of geometrical singularity. In fig.\ref{f2}, we have shown the variation of the energy density($\rho$), radial pressure ($p_r$) and tangential pressure ($p_t$) respectively. It is clearly visible that all these three parameters are maximum at the center and monotonically decreasing towards the surface so that one can achieve minimum result on the surface. Fig.\ref{f3} depicts the anisotropy feature of the system that shows minimum value of anisotropy i.e zero at the center and maximum at the boundary in the present modified $f(\Re,T)$ gravity theory where in general relativity maximum anisotropy at the boundary carries the property of the anisotropic strange stars.
		
		\par Also in order to evaluate the physical liability of this stellar model in the framework of $f(\Re,T)$ gravity theory, we have examined the energy conditions, in fig.\ref{f4} which tells that our stellar structure is consistent with all the energy conditions. Then in fig.\ref{f5}, we presented the variation in compactness factor and in fig.\ref{f6}, we shown the gravitational and surface redshift of the compact star Her X-1. The variation in the casuality conditions and Herrera-Cracking concept are studied and presented against radial coordinate $r/R$ in figs.\ref{f7} and \ref{f8}, respectively. Further in fig.\ref{f9}, we presented the variation of the adiabatic index against radial coordinate $r/R$ and concluded that the value of $\Gamma$ is greater than $4/3$ which confirms that our model is stable against the radial pulsation. The variation of EoS parameters as in fig.\ref{f10}, depicts that the anisotropic fluid distribution is real and naturaly non-exotic.
		
		\par Lastly to show the stability of the stellar model in terms of the equilibrium of forces we have examined the modified TOV equation as in fig.\ref{f11} in the background of $f(\Re,T)$ theory of gravity. The variation of all four forces namely gravitational force, hydrostatic force, anisotropic force and force due to modified gravity has been shown for different values of $\xi$. We see that the gravitational force ($F_g$) is always negative  and hydrostatic force ($F_h$) is always positive, the anisotropic force ($F_a$) is almost negligible compared to other forces so that it coincides along X-axis. Whereas the modified force ($F_m$) is in between the hydrostatic force and anisotropic force. That concludes that $F_g$ is balanced by both $F_h$ and $F_m$ for our stellar structure. In fig.\ref{f12}, we have presented the varialtion of bag constant against coupling constant ($\xi$) which shows that the values of the bag constant are less than the value of $\xi$ i.e $\xi=0, 0.2, 0.4, 0.6, 0.8, 1$. In table \ref{T1}, we obtained central density, surface density, central radial pressure, adiabatic index, surface redshift, compactness and bag constant for different values of $\xi$ for the compact star Her X-1 by assuming $M=0.85M_{\odot}$ and $R=8.1km$.  
		
		\par Einstein's theory has been successfully tested mainly in the regime of weak gravity through solar system and laboratory experiments. But still this theory faces strong stringent constraints in the region of strong gravity specially near the region of black hole, highly densed compact stars and expanding Universe. It has been noticed that because of $f(\Re,T)$ theory, the maximal mass limits rise higher than their standard values in general relativity for different parametric values of $\xi$. Hence, such stellar configurations in the background of modified theory gravity can also be used in explaining massive stellar systems like massive pulsars, super-Chandrasekhar stars and magnetstars,etc which are not clearly explained by general Einstein's theory. Now, to summarize the paper, we have obtained a well-behaved model for anisotropic compact stars in presence of $f(\Re,T)$ modified theory of gravity and its results have been analyzed both analytically and graphically. We have shown the solution of stellar system depends on the MIT bag model EoS which is already used by several researchers for modeling compact stars. Finally, it is important to say that by taking $\xi\implies 0$, results of GR theory in four dimensions are recovered.

\section*{Acknowledgment}
The Authors would like to express their sincere gratitude towards Department of Mathematics, Central University of Jharkhand, Ranchi, India for the necessary support and encouragement while writing this paper and finalizing.

\end{document}